\documentstyle[12pt]{article}
\title{Vacuum in quantum liquids and in general relativity}
\author{ G.E. Volovik\\
Low Temperature Laboratory,
Helsinki University of Technology\\
P.O.Box 2200, FIN-02015 HUT, Finland\\
and\\
L.D. Landau Institute for Theoretical Physics, \\
Kosygin Str. 2, 117940 Moscow, Russia\\
}
\begin{document}

\maketitle
\vskip 2 truecm

\begin{abstract}
{Quantum liquids, in which an effective Lorentzian metric
and thus some kind of gravity  gradually arise in the
low-energy corner, are the objects where the problems
related to the quantum vacuum can be investigated in
detail. In particular, they provide the possible solution
of the cosmological constant problem: why the vacuum
energy is by 120 orders of magnitude smaller than the
estimation from the relativistic quantum field theory.
The almost complete cancellation of the cosmological
constant does not require any fine tuning and comes from
the fundamental ``trans-Planckian'' physics of quantum
liquids. The remaining vacuum energy is generated by the
perturbation of quantum vacuum caused by matter
(quasiparticles), curvature, and other possible sources,
such as smooth component -- the quintessence. This
provides the possible solution of another cosmological
constant problem: why the present cosmological constant is
on the order of the present matter density of the Universe.
We discuss here some properties of the quantum vacuum in
quantum liquids: the vacuum energy under different
conditions; excitations above the vacuum state and the
effective acoustic metric for them provided by the motion
of the vacuum; Casimir effect, etc.
}

\end{abstract}

\tableofcontents

\section{Introduction.}

Quantum liquids, such as $^3$He and $^4$He, represent the
systems of strongly interacting and strongly
correlated atoms, $^3$He and $^4$He atoms correspondingly.
Even in its ground state, such liquid is a rather
complicated object, whose many body physics requires
extensive numerical simulations. However, when the energy
scale is reduced below about 1 K, we cannot resolve anymore
the motion of isolated atoms in the liquid. The
smaller the energy the better is the liquid described in
terms of the collective modes and the dilute gas of
the particle-like excitations -- quasiparticles. This is
the Landau picture of the low-energy degrees of
freedom in quantum Bose and Fermi liquids. The dynamics of
collective modes and quasiparticles is described in terms
of what we call now `the effective theory'. In superfluid
$^4$He this effective theory, which incorporates the
collective motion of the ground state -- the quantum
vacuum -- and the dynamics of quasiparticles in the
background of the moving vacuum, is known as the two-fluid
hydrodynamics
\cite{Khalatnikov}.

Such an effective theory does not
depend on details of microscopic (atomic) structure of
the quantum liquid. The type of the effective theory is
determined by the symmetry and topology of the ground
state, and the role of the microscopic physics is only to
choose between different universality classes on the basis
of the minimum energy consideration. Once the universality
class is determined, the low-energy properties of the
condensed matter system are completely described by the
effective theory, and the information on the microscopic
(trans-Planckian) physics is lost
\cite{LaughlinPines}.

In some condensed matter the universality class
produces the effective theory, which reminds very closely
the relativistic quantum field theory. For example, the
collective fermionic and bosonic modes in superfluid
$^3$He-A reproduce chiral fermions, gauge fields and
even in many respects the gravitational field
\cite{PhysRepRev}.

This allows us to use the quantum
liquids for the investigation of the properties related to
the quantum vacuum in relativistic quantum field theories,
including the theory of gravitation.  The main advantage
of the quantum liquids  is that in principle we
know their vacuum structure at any relevant scale,
including the interatomic distance,  which plays the part
of one of the Planck length scales in the hierarchy of
scales. Thus the quantum liquids can provide possible
routes from our present low-energy corner
of the effective theory  to the ``microscopic'' physics  at
Planckian and trans-Planckian energies.

One of the possible routes is related to the conserved
number of atoms $N$ in the quantum liquid. The quantum
vacuum of the quantum liquids is constructed from the
discrete elements, the bare atoms. The interaction and
zero point motion of these atoms compete and provide
an equilibrium ground state of the ensemble of atoms,
which can exist even in the absence of external pressure.
The relevant energy and the pressure in this equilibrium
ground state are exactly zero in the absence of
interaction with environment. Translating this to the
language of general relativity, one obtains that the
cosmological constant in the effective theory of gravity
in the quantum liquid is exactly zero without any fine
tuning. The equilibrium quantum vacuum is not gravitating.

This route shows a possible solution of the cosmological
constant problem: why the estimation the vacuum energy
using the relativistic quantum field theory gives the
value, which is by 120 orders of magnitude higher than its
upper experimental limit. In quantum liquids there is
a similar discrepancy between the exact zero result for
the vacuum energy and the naive estimation within the
effective theory.  We shall also discuss here how different
perturbations of the vacuum in quantum liquids lead to
small nonzero energy of quantum vacuum. Translating this
to the language of general relativity, one obtains that
the in each epoch the vacuum energy density must be
either of order of the matter density of the Universe,
or of its curvature, or of the energy density of
the smooth component -- the quintessence.

Here we mostly discuss the Bose ensemble of atoms: a
weakly interacting Bose gas, which experiences the
phenomenon of Bose condensation, and a real Bose liquid --
superfluid $^4$He. The consideration of the Bose gas allows
us to use the microscopic theory to derive the ground
state energy of the quantum system of interacting atoms and
the excitations above the vacuum state -- quasiparticles.
We also discuss the main differences between the bare
atoms, which comprise the vacuum state, and the
quasiparticles, which serve as elementary particles in the
effective quantum field theory.

Another consequence of the discrete
number of the elements comprising the vacuum state, which
we consider, is related to the Casimir effects. The
dicreteness of the vacuum -- the finite-$N$
effect -- leads to the to the mesoscopics Casimir forces,
which cannot be derived within the effective theory. For
these perposes we consider the Fermi ensembles of atoms:
Fermi gas and Fermi liquid.

\section{Einstein gravity and cosmological constant
problem}

\subsection{Einstein action}

The Einstein's geometrical theory of gravity  consists of
two main elements \cite{Damour}.

(1)  Gravity is related to a curvature of space-time  in
which particles move along the geodesic curves in the
absence of non-gravitational forces. The geometry of the
space-time  is described by the metric  $g_{\mu\nu}$ which
is the dynamical field of gravity. The action for matter
in the presence of gravitational field $S_{\rm M}$, which
simultaneously describes the coupling between gravity and
all other fields  (the matter fields), is obtained from
the special relativity action for the matter fields  by
replacing everywhere the flat Minkowski metric by the
dynamical metric
$g_{\mu\nu}$ and the partial derivative by
$g$-covariant derivative. This follows from  the principle
that the equations of motion do not depend on the choice
of the coordinate system (the so called general
covariance). This also means that the motion in the
non-inertial frame can be described in the same manner as
motion in some gravitational field -- this is the
equivalence principle. Another consequence of the
equivalence principle is that the the space-time geometry
is the same for all the particles: the gravity is
universal.

(2) The dynamics of the gravitational field is  determined
by adding the action functional $S_{\rm G}$ for
$g_{\mu\nu}$, which describes propagation and
self-interaction of the gravitational field:
\begin{equation}
S=S_{\rm G}+S_{\rm M}~.
\label{GravitationalEinsteinAction}
\end{equation}
The general covariance requires  that $S_{\rm G}$ is the
functional of the curvature.  In the original Einstein
theory only the first order curvature term is retained:
\begin{equation}
S_{\rm G}= -{1\over
16\pi G}\int d^4x\sqrt{-g}{\cal R}~,
\label{EinsteinCurvatureAction}
\end{equation}
where $G$ is gravitational Newton cosntant;  and ${\cal
R}$ is the Ricci scalar curvature. The Einstein action is
thus
\begin{equation}
S= -{1\over
16\pi G}\int d^4x\sqrt{-g}{\cal R} +S_{\rm M}
\label{EinsteinAction}
\end{equation}
Variation of this action over the metric  field
$g_{\mu\nu}$ gives the Einstein equations:
\begin{equation}
{\delta S\over \delta g^{\mu\nu}}= {1\over
2}\sqrt{-g}\left[ -{1\over 8\pi G}\left(
R_{\mu\nu}-{1\over 2}{\cal R}g_{\mu\nu} \right)  +T^{\rm
M}_{\mu\nu}\right]=0~,
\label{EinsteinEquation1}
\end{equation}
where $T^{\rm M}_{\mu\nu}$ is the energy-momentum  of the
matter fields. Bianchi identities lead to the
``covariant'' conservation law for matter
\begin{equation}
T^{\mu {\rm M}}_{\nu;\mu}=0 ~~,~~{\rm or}~~
\partial_\mu \left(T^{\mu {\rm M}}_\nu \sqrt{-g}\right) =
{1\over 2}\sqrt{-g}T^{\alpha\beta {\rm M}} \partial_\nu
g_{\alpha\beta}
 \, ,
\label{CovariantConservation}
\end{equation}
But actually  this ``covariant'' conservation takes place
in virtue of the field equation for ``matter'' irrespective
of the dynamics of the gravitational field.

\subsection{Vacuum energy and cosmological term}

In particle physics, field quantization allows a zero point energy, the
constant
energy when all fields are in their ground states. In the absence of gravity,
only the difference between zero points can be measured, for example in the
Casimir effect, while the absolute value in unmeasurable, However, Einstein's
equations react to $T^{\rm M}_{\mu\nu}$ and thus to the value of vacuum energy
itself.

If the vacuum energy is taken seriously, the energy-momentum tensor of the
vacuum
must be added to the Einstein equations. The corresponding action is given by
the so-called cosmological term, which was introduced by Einstein in 1917
\cite{EinsteinCosmCon}:
\begin{equation}
S_\Lambda = -\rho_\Lambda
\int d^4x\sqrt{-g}~,~~T^\Lambda_{\mu\nu}={2\over \sqrt{-g}}{\delta
S_\Lambda\over
\delta g^{\mu\nu}}=\rho_\Lambda g_{\mu\nu}~.
\label{CosmologicalTerm}
\end{equation}
The energy-momentum tensor of the vacuum shows that the quantity
$\rho_\Lambda\sqrt{-g}$ is the vacuum energy density, and the equation of
state of
the vacuum is
\begin{equation}
P_\Lambda=-\rho_\Lambda~,
\label{EquationOfState}
\end{equation}
where $P_\Lambda\sqrt{-g}$ is the
partial pressure of the vacuum. The Einstein's equations are modified:
\begin{equation}
 {1\over 8\pi G}\left(
R_{\mu\nu}-{1\over 2}{\cal R}g_{\mu\nu} \right) = T^\Lambda_{\mu\nu}
 +T^{\rm M}_{\mu\nu}~.
\label{EinsteinEquation}
\end{equation}

\subsection{Cosmological constant problem}

The most severe problem in the marriage of gravity and quantum theory is why
is the vacuum not gravitating \cite{WeinbergReview}. The
vacuum energy density can be easily estimated: the
positive contribution comes from the zero-point energy of
the bosonic fields and the negative -- from the occupied
negative energy levels in the Dirac sea. Since the largest
contribution comes from the high momenta, where the energy
spectrum of particles is massless, $E=cp$, the energy
density of the vacuum is
\begin{equation}
\rho_\Lambda \sqrt{-g}= {1\over V}\left(\nu_{\rm bosons}\sum_{\bf p}{1\over 2}
cp~~ -\nu_{\rm fermions}
\sum_{\bf p} cp\right)
\label{VacuumEnergyPlanck}
\end{equation}
where $V$ is the volume of the system; $\nu_{\rm bosons}$ is the number of
bosonic species and
$\nu_{\rm fermions}$ is the number of fermionic species. The vacuum energy is
divergent and the natural cut-off is provided by the gravity itself. The
cut-off
Planck energy is determined by the Newton's constant:
\begin{equation}
E_{\rm Planck}=\left({\hbar  c^5\over G}\right)^{1/2}~,
\label{PlanckCutoff}
\end{equation}
It is on the order of $10^{19}$ GeV. If there is no symmetry between the
fermions
and bosons (supersymmetry) the Planck energy scale cut-off provides the
following
estimation for the vacuum energy density:
\begin{equation}
\rho_\Lambda \sqrt{-g}\sim
 \pm {1\over c^3} E^4_{\rm Planck} =\pm \sqrt{-g}~E^4_{\rm Planck}~,
\label{VacuumEnergyPlanck2}
\end{equation}
with the sign of the vacuum energy being
determined by the fermionic and bosonic content of the quantum field
theory. Here we
considered the flat space with Minkowski metric $g_{\mu\nu}={\rm
diag}(-1,c^{-2},c^{-2},c^{-2})$.

The ``cosmological constant problem'' is a huge
disparity,  between the naively expected value in
Eq.(\ref{VacuumEnergyPlanck2}) and the range of actual values:
the experimental observations show that
$\rho_\Lambda$ is less than or on the order of $10^{-120} E^4_{\rm Planck}$
\cite{Supernovae}. In case of supersymmetry, the cut-off is somewhat less,
being
determined by the scale at which supersymmetry is violated, but the huge
disparity
persists.

This disparity demonstrates that the vacuum energy  in
Eq.~(\ref{VacuumEnergyPlanck}) is not gravitating.  This
is in apparent contradiction with the general principle of
equivalence, according to which the inertial and
gravitating masses must coincide. This indicate that the
theoretical criteria for setting the absolute zero point
of energy are unclear and probably require the physics
beyond general relativity. To clarify this issue we can
consider such quantum systems where the elements of the
gravitation are at least partially reproduced, but where
the structure of the quantum vacuum is known. Quantum
liquids are the right systems.

\subsection{Sakharov induced gravity}

Why is the Planck energy  in Eq.(\ref{PlanckCutoff}) the natural cutoff in
quantum
field theory? This is based on the important observation made by
Sakharov that the second element of the Einstein's theory
can follow from the first one due to the quantum
fluctuations of the relativistic matter field
\cite{Sakharov}. He showed that vacuum fluctuations of the matter field
induce the
curvature term in action for $g_{\mu\nu}$. One can even argue that the whole
Einstein action is induced by vacuum polarization, and thus the gravity is
not the
fundamental force but is determined by the properties of the quantum vacuum.

The magnitude of the induced Newton's constant is
determined by the value of the cut-off: $G^{-1}\sim \hbar
E^2_{\rm cutoff}/c^5$. Thus in this  Sakharov's gravity
induced by quantum fluctuations the causal connection
between the gravity and the cut-off is reversed: the
physical high-energy cut-off determines the gravitational
constant.  The $E^2_{\rm cutoff}$ dependence of the
inverse gravitational constant explains why the gravity is
so small compared to the other forces, whose running
coupling ``constants'' have only mild logarithmic
dependence on $E_{\rm cutoff}$.

The same cut-off must be applied for the estimation of the
cosmological constant, which thus must be of order of
$E^4_{\rm cutoff}$. But this is in severe contradiction
with experiment. This shows that, though the effective
theory is apprpopriate for the calculation of the Einstein
curvature term, it is not applicable for the calculation
of the vacuum energy: the trans-Planckian physics must be
evoked for that.

The Sakharov  theory does not explain the
first element of the Einstein's theory: it does not show
how the metric field $g_{\mu\nu}$ appears. This can be
given only by the fundamental theory of quantum vacuum,
such as string theory where the gravity appears as a
low-energy mode. The quantum liquid examples also show
that the metric field can naturally and in some cases even
emergently appear as the low-energy collective mode of the
quantum vacuum.

\subsection{Effective gravity in quantum liquids}

The first element of the Einstein theory of gravity (that the motion of
quasiparticles is governed by the effective curved space-time) arises in
many condensed matter systems in the low-energy limit. An example is the
motion of
acoustic phonons in distorted crystal lattice, or in the background flow
field of
superfluid condensate.  This motion is described by the
effective acoustic metric
\cite{Unruh,Vissersonic,StoneIordanskii,StoneMomentum}.
For this ``relativistic matter field'' (acoustic phonons
with dispersion relation $E=cp$, where $c$ is a speed of
sound, simulate relativistic particles) the analog of the
equivalence principle is fulfilled. As a result the
covariant conservation law in
Eq.(\ref{CovariantConservation}) does hold for the
acoustic mode, if
$g_{\mu\nu}$ is replaced by the acoustic metric.

The second element of the Einstein's gravity is not
easily reproduced in condensed matter. In general, the
dynamics of acoustic metric $g_{\mu\nu}$ does not obey the
equivalence principle inspite of the Sakharov mechanism of
the induced gravity. In the existing quantum liquids the
Einstein action induced by the quantum fluctuations of the
``relativistic matter field'' is much smaller than the
non-covariant action induced by ``non-relativistic''
high-energy component of the quantum vacuum, which is
overwhelming in these liquids.  Of course, one can find
some very special cases  where the Einstein action for the
effective metric is dominating, but this is not a rule.

Nevertheless, inspite of the incomlete analogy with the
Einstein theory, the effective gravity in quantum liquids
can be useful for investigation of the cosmological
constant problem.

\section{Microscopic `Theory of Everything' in quantum
liquids}

\subsection{Microscopic and effective theories}

There are two ways to study quantum liquids:

(i) The fully microscopic treatment. It can be realized
completely  (a) by numerical simulations of the many
body problem; (b) analytically for some special models;
(3) perturbatively for some special ranges of the material
parameters, for example, in the limit of weak interaction
between the particles.

(ii) Phenomenological approach in terms of effective
theories. The hierarchy of the effective theories
correspond to the low-frequency long-wave-length dynamics
of quantum liquids in different ranges of frequency.
Examples of effective theories: Landau theory of Fermi
liquid; Landau-Khalatnikov two-fluid hydrodynamics of
superfluid $^4$He \cite{Khalatnikov}; theory of elasticity
in solids; Landau-Lifshitz theory of ferro- and
antiferromagnetism; London theory of superconductivity;
Leggett theory of spin dynamics in superfluid phases of
$^3$He; effective quantum electrodynamics arising in
superfluid $^3$He-A; etc. The latter example indicates,
that the existing Standard Model of electroweak, and
strong interactions, and the Einstein gravity too, are the
phenomenological effective theories of high-energy
physics, which describe its low-energy edge, while the
microscopic theory of the quantum vacuum is absent.

\subsection{Theory of Everything for quantum liquid}

The microscopic ``Theory of Everything'' for quantum
liquids -- ``a set of equations capable of describing all
phenomena that have been observed''
\cite{LaughlinPines} in these quantum systems --  is
extremely simple. On the ``fundamental'' level appropriate
for quantum liquids and solids, i.e. for all practical
purposes, the $^4$He or $^3$He atoms of these quantum
systems can be considered as structureless: the $^4$He
atoms are the structureless bosons and the  $^3$He atoms
are the structureless fermions with spin $1/2$. The Theory
of Everything for a collection of a macroscopic number $N$
of interacting $^4$He or $^3$He atoms is contained in the
non-relativistic many-body Hamiltonian
\begin{equation}
{\cal H}= -{\hbar^2\over 2m}\sum_{i=1}^N {\partial^2\over
\partial{\bf r}_i^2} +\sum_{i=1}^N\sum_{j=i+1}^N U({\bf
r}_i-{\bf r}_j)~,
\label{TheoryOfEverythingOrdinary}
\end{equation}
acting on the many-body wave function  $\Psi({\bf
r}_1,{\bf r}_2, ... , {\bf r}_i,... ,{\bf r}_j,...)$. Here
$m$ is the bare mass of the atom; $U({\bf r}_i-{\bf r}_j)$
is the pair  interaction of the bare atoms $i$ and $j$.
When written in the second quantized form it becomes the
Hamiltonian of the quantum field theory
\begin{equation}
{\cal H}-\mu {\cal N}=\int d{\bf x}\psi^\dagger({\bf
x})\left[-{\nabla^2\over 2m}
-\mu
\right]\psi({\bf x}) +{1\over 2}\int d{\bf x}d{\bf y}U({\bf x}-{\bf
y})\psi^\dagger({\bf x})
\psi^\dagger({\bf y})\psi({\bf y})\psi({\bf x})
\label{TheoryOfEverything}
\end{equation}
In $^4$He, the bosonic quantum field
$\psi({\bf x})$ is  the annihilation operator of the
$^4$He atoms. In $^3$He,
$\psi({\bf x})$ is the fermionic field  and the spin
indices must be added.  Here  ${\cal N}=\int d{\bf
x}~\psi^\dagger({\bf x})\psi({\bf x})$ is the operator of particle number
(number of
atoms);
$\mu$ is the chemical potential --  the Lagrange
multiplier which is introduced to take into account the
conservation of the number of atoms.

\subsection{Importance of discrete particle number in
microscopic theory}

This is the main
difference from the relativistic quantum  field theory,
where the number of any particles is not restricted:
particles and antiparticles can be created from the
quantum vacuum. As for the number of particles in the
quantum vacuum itself, this quantity is simply not
determined today. At the moment we do not know the
structure of the quantum vacuum and its particle content.
Moreover, it is not clear whether it is possible to
describe the vacuum in terms of some discrete elements
(bare particles) whose number is conserved. On the
contrary, in quantum liquids the analog of the quantum
vacuum -- the ground state of the quantum liquid -- has
the known number of atoms. If
$N$ is big, this difference between the two quantum  field
theories disappears. Nevertheless, the mear fact, that
there is a conservation law for the number of particles
comprising the vacuum, leads to the definite conclusion on
the value of the relevant vacuum energy.  Also, as we
shall see below in Sec. \ref{EffectsDiscreteNumberN}, the
discreteness of the quantum vacuum can be revealed in the
mesoscopic Casimir effect.

\subsection{Enhancement of symmetry in the low energy
corner. Appearance of effective theory.}

The Hamiltonian (\ref{TheoryOfEverything}) has  very
restricted number of symmetries: It is invariant under
translations and $SO(3)$ rotations in 3D space;  there is
a global $U(1)$ group originating from the conservation of
the number of atoms: ${\cal H}$ is invariant under gauge
rotation $\psi({\bf x})\rightarrow e^{i\alpha}\psi({\bf
x})$ with constant $\alpha$; in $^3$He in addition, if the
relatively weak spin-orbit coupling is neglected, ${\cal
H}$ is also invariant under separate rotations of spins,
$SO(3)_S$. At low temperature the phase transition to the
superfluid or to the quantum crystal state occurs where
some of these symmetries are broken spontaneously. For
example, in the $^3$He-A state  all of these symmetries,
except for the translational symmetry, are broken.

However, when the temperature and energy decrease further
the symmetry becomes gradually enhanced in agreement with
the anti-grand-unification scenario
\cite{FrogNielBook,Chadha}.  At low energy the quantum
liquid or solid is well described in terms of a dilute
system of quasiparticles. These are bosons (phonons) in
$^4$He and fermions and bosons in
$^3$He, which move in the background of the effective
gauge and/or gravity fields simulated by the dynamics of
the collective modes. In particular, phonons propagating
in the inhomogeneous liquid are described by the effective
Lagrangian
\begin{equation}
L_{\rm effective}=\sqrt{-g}g^{\mu\nu}\partial_\mu\alpha
\partial_\nu\alpha~,
\label{LagrangianSoundWaves}
\end{equation}
where $g^{\mu\nu}$ is the effective acoustic  metric
provided by inhomogeneity and flow of the liquid
\cite{Unruh,Vissersonic,StoneMomentum}.

These quasiparticles serve as the elementary particles  of
the low-energy effective quantum field theory. They
represent the analogue of matter. The type of the
effective quantum field theory -- the theory of
interacting fermionic and bosonic quantum fields --
depends on the universality class of the fermionic
condensed matter (see review \cite{PhysRepRev}).  The
superfluid $^3$He-A, for example, belongs to the same
universality class as the Standard Model. The effective
quantum field theory describing the low energy
phenomena in  $^3$He-A, contains chiral ``relativistic''
fermions. The collective bosonic modes interact with
these ``elementary particles'' as gauge fields and
gravity. All these fields emergently arise together with
the Lorentz and gauge invariances and with elements of the
general covariance from the fermionic Theory of Everything
in Eq.(\ref{TheoryOfEverything}).

The emergent phenomena do not depend much on  the details
of the Theory of Everything \cite{LaughlinPines}, in our
case on the details of the pair potential
$U({\bf x}-{\bf y})$. Of course, the latter determines the universality
class in
which the system enters at low energy.  But once the
universality class is established, the physics remains
robust to deformations of the pair potential. The details
of $U({\bf x}-{\bf y})$ influence only the ``fundamental''
parameters of the effective theory (``speed of light'',
``Planck'' energy cut-off, etc.) but not the general
structure of the theory. Within the effective theory the
``fundamental'' parameters are considered as
phenomenological.

\section{Weakly interacting Bose gas}

The quantum liquids are strongly correlated and  strongly
interacting systems. That is why, though it is possible to
derive the effective theory from first principles in
Eq.(\ref{TheoryOfEverything}), if one has enough computer
time and memory, this is a rather difficult task. It is
instructive, however, to consider the microscopic theory
for some special model potentials $U({\bf x}-{\bf y})$.
This allow us to solve the problem completely or
perturbatively. In case of the Bose-liquids the proper
model is the Bogoliubov weakly interacting Bose gas, which
is in the same universality class as a real superfluid
$^4$He.  Such model is very useful, since it
simultaneously  covers the low-energy edge of the effective
theory, and the high-energy ``transPlanckian'' physics.

\subsection{Model Hamiltonian}

Here we follow mostly the book by Khalatnikov
\cite{Khalatnikov}.  In the model of weakly interacting
Bose gas the pair potential in
Eq.(\ref{TheoryOfEverything}) is weak.   As a result the
most of the particles at $T=0$ are in the Bose condensate,
i.e. in the state with the momentum ${\bf p}=0$. The Bose
condensate is characterized by the nonzero vacuum
expectation value (vev) of the particle annihilation
operator at ${\bf p}=0$:
\begin{equation}
\left<a_{{\bf p}=0}\right>=\sqrt{N_0}e^{i\Phi}~, ~\left<a^\dagger_{{\bf
p}=0}\right>=\sqrt{N_0}e^{-i\Phi}.
\label{VEV}
\end{equation}
Here $N_0$ is the particle number  in the Bose condensate,
and $\Phi$ is the phase of the condensate. The vacuum is
degenerate over global $U(1)$ rotation of the phase.
Further we consider particular vacuum state with $\Phi=0$.

If there is no interaction between  the particles (an
ideal Bose gas), all the particles at $T=0$ are in the
Bose condensate, $N_0=N$. Small interaction induces a
small fraction of particles which are not in condensate,
these particle have small momenta
${\bf p}$. As a result only zero Fourier component  of the
pair potential is relevant, and
Eq.(\ref{TheoryOfEverything}) has the form:
\begin{eqnarray}
{\cal H}-\mu {\cal N}=-\mu N_0+{N_0^2U\over 2V}
+\label{TheoryOfEverythingBoseGas1}\\
\sum_{{\bf p}\neq 0}\left({p^2\over
2m}- \mu\right)a^\dagger_{\bf p}a_{\bf p}+{N_0U\over
2V}\sum_{{\bf p}\neq 0}\left(2a^\dagger_{\bf p}a_{\bf
p}+2a^\dagger_{-{\bf p}}a_{-{\bf p}}+a_{{\bf p}}a_{-{\bf
p}}+a^\dagger_{{\bf p}}a^\dagger_{-{\bf p}}\right)
\label{TheoryOfEverythingBoseGas2}
\end{eqnarray}
Here
$N_0=a^\dagger_0a_0=a_0a^\dagger_0=
a^\dagger_0a^\dagger_0=a_0a_0$
is the particle number in the Bose-condensate (we
neglected quantum fluctuations of the operator $a_0$ and
consider $a_0$ as $c$-number);
$U$ is the matrix element of pair interaction for  zero
momenta ${\bf p}$ of particles. Minimization of the main
part of the energy in
Eq.(\ref{TheoryOfEverythingBoseGas1}) over $N_0$ gives
$UN_0/V=\mu$ and one obtains:
\begin{eqnarray}
{\cal H}-\mu {\cal N}=-{\mu^2 \over 2U}V
+\sum_{{\bf p}\neq 0}{\cal H}_{\bf p}
\label{TheoryOfEverythingBoseGas3}\\ {\cal H}_{\bf p} =
{1\over 2}\left({p^2\over 2m}+\mu\right)
\left(a^\dagger_{\bf p}a_{\bf p}+a^\dagger_{-{\bf
p}}a_{-{\bf p}}\right)+{\mu\over 2}\left( a_{{\bf
p}}a_{-{\bf p}}+a^\dagger_{{\bf p}}a^\dagger_{-{\bf
p}}\right)
\label{TheoryOfEverythingBoseGas4}
\end{eqnarray}

\subsection{Pseudorotation --
Bogoliubov transformation}

At each ${\bf p}$ the Hamiltonian can be diagonalized
using the following consideration. The operators
\begin{equation}
{\cal L}_3={1\over 2}(a^\dagger_{{\bf p}} a_{{\bf p}}+
a^\dagger_{-{\bf p}} a_{-{\bf p}}+1)~~,~~  {\cal L}_1+i
{\cal L}_2= a^\dagger_{{\bf p}} a^\dagger_{-{\bf p}}~~,~~
{\cal L}_1-i {\cal L}_2=a_{{\bf p}} a_{-{\bf p}}
\label{PseudoRotationOperators}
\end{equation}
form the group of pseudorotations, $SU(1,1)$  (the group
which conserves the form
$x_1^2+x_2^2-x_3^2$), with the commutation relations:
\begin{equation}
[{\cal L}_3, {\cal L}_1]=i {\cal L}_2~,~[ {\cal L}_2,{\cal L}_3]=i {\cal
L}_1~,~[
{\cal L}_1, {\cal L}_2]=-i{\cal L}_3~,
\label{PseudoRotationOperatorsCommutations}
\end{equation}
In terms of the pseudomomentum the Hamiltonian in
Eq.(\ref{TheoryOfEverythingBoseGas4}) has the form
\begin{equation}
{\cal H}_{\bf p} = \left({p^2\over
2m}+\mu\right){\cal L}_3 +\mu {\cal L}_1- {1\over 2}\left({p^2\over
2m}+\mu\right)~.
\label{TheoryOfEverythingBoseGas5}
\end{equation}
In case of the nonzero phase $\Phi$ of the Bose condensate
one has
\begin{equation}
{\cal H}_{\bf p} = \left({p^2\over
2m}+\mu\right){\cal L}_3 +\mu\cos(2\Phi)~{\cal
L}_1+\mu\sin(2\Phi)~ {\cal L}_2- {1\over 2}\left({p^2\over
2m}+\mu\right)~.
\label{BoseGasNonZeroPhi}
\end{equation}

The diagonalization of this Hamiltoiniaby  is achieved
first by rotation by angle $2\Phi$ around axis $z$, and
then by the Lorentz transformation -- pseudorotation
around axis $y$:
\begin{equation}
{\cal L}_3=\tilde {\cal L}_3{\rm ch}\chi + \tilde {\cal L}_1{\rm sh}\chi ~~,~~
{\cal L}_1 =\tilde  {\cal L}_1{\rm ch}\chi +\tilde {\cal L}_3{\rm sh}\chi_y
~~,~~{\rm th}\chi  ={\mu\over {p^2\over
2m}+\mu}~~.
\label{Pseudorotation}
\end{equation}
This corresponds to Bogoliubov transformation  and gives
the following diagonal Hamiltonian:
\begin{eqnarray}
{\cal H}_{\bf p} =- {1\over 2}\left({p^2\over
2m}+\mu\right)+\tilde {\cal L}_3 \sqrt{\left({p^2\over
2m}+\mu\right)^2 -\mu^2}=
\label{DiagonalizedHamiltonian1}\\
= {1\over 2}  E({\bf p})\left(\tilde a^\dagger_{\bf p}\tilde a_{\bf p}+
\tilde a^\dagger_{-{\bf p}}\tilde a_{-{\bf p}}\right)
+ {1\over
2}
\left(E({\bf p}) -\left({p^2\over 2m}+\mu\right) \right)~,
\label{DiagonalizedHamiltonian2}
\end{eqnarray}
where $\tilde a_{\bf p}$ is the operator of  annihilation
of quasiparticles, whose energy spectrum $E({\bf p})$ is
\begin{equation}
E({\bf p})=\sqrt{\left({p^2\over
2m}+\mu\right)^2 -\mu^2}=\sqrt{p^2c^2 +{p^4\over
4m^2}}~~,~~c^2={\mu\over m}~.
\label{BogoliubovEnergySpectrum}
\end{equation}

\subsection{Vacuum and quasiparticles}

 The total Hamiltonian now represents the ground state  --
the vacuum -- and the system of quasiparticles
\begin{equation}
{\cal H}-\mu {\cal N}=\left<{\cal H}-\mu {\cal N}\right>_{\rm vac} +
\sum_{\bf p}
E({\bf p})\tilde a^\dagger_{\bf p}\tilde a_{\bf p}
\label{EnergyBoseGas}
\end{equation}
The lower the energy the more dilute is  the system of
quasiparticles and thus the
 weaker is the interaction between them.  This description
in terms of the vacuum state and dilute system of
quasiparticle is generic for the condensed matter systems
and is valid even if the interaction of the initial bare
particles is strong. The phenomenological effective theory
in terms of vacuum state and quasiparticles was developed
by Landau both for Bose and Fermi liquids.  Quasiparticles
(not the bare particles) play the role of elementary
particles in such effective quantum field theories.

In a weakly interacting Bose-gas
in Eq.(\ref{BogoliubovEnergySpectrum}), the spectrum of
quasiparticles at low energy (i.e. at  $p\ll mc$) is
linear,
$E=cp$.  The linear slope coincides with the speed  of
sound, which can be obtained from the leading term in
energy: $N(\mu)=-d(E-\mu N)/d\mu=\mu V/U$,
$c^2=N(d\mu/dN)/m=\mu/m$. These quasiparticles are
phonons -- quanta of sound waves. The same quasiparticle
spectrum occurs in the real superfluid liquid
$^4$He, where the interaction between the bare particle
is strong. This shows that the qualitative low-energy
properties of the system do not depend on the
microscopic (trans-Planckian) physics. The latter
determines only the speed of sound $c$. One can say, that
weakly and strongly interacting Bose systems belong to the
same universality class, and thus have the same low-energy
properties. One cannot distinguish between the two systems
if the observer measures only the low-energy effects,
since they are described by the same effective theory.

\subsection{Particles and quasiparticles}

It is necessary to distinguish
between the bare particles and quasiparticles.

Particles are the elementary objects of the system on a
microscopic ``transPlanckian'' level, these are the atoms
of the underlying liquid ($^3$He or $^4$He atoms). The
many-body system of the interacting atoms form the quantum
vacuum -- the ground state. The nondissipative collective
motion of the superfluid vacuum with zero entropy is
determined by the conservation laws experienced by the
atoms and by their quantum coherence in the superfluid
state.

Quasiparticles are the particle-like
excitations above this vacuum state, they serve as
elementary particles in the effective theory. The bosonic
excitations in superfluid $^4$He and fermionic and bosonic
excitations in superfluid $^3$He represent  the matter in
our analogy.  In superfluids they form the viscous normal
component responsible for the thermal and kinetic
low-energy properties of superfluids. Fermionic
quasiparticles in $^3$He-A are chiral fermions, which are
the counterpart of the leprons and quarks in the Standard
Model \cite{PhysRepRev}.

\subsection{Galilean
transformation for particles and quasiparticles}
\label{GalileanTransformation}

The quantum liquids considered  here are essentially
nonrelativistic: under the laboratory conditions their
velocity is much less than the speed of light. That is
why they obey with great precision the Galilean
transformation law. Under the Galilean transformation to
the coordinate system moving with the velocity ${\bf u}$
the superfluid velocity -- the velocity of the quantum
vacuum -- transforms as ${\bf v}_{\rm s}\rightarrow {\bf
v}_{\rm s} + {\bf u}$.

As for the transformational  properties of bare particles
(atoms) and   quasiparticles, it appears that they are
essentially  different. Let us start with bare particles.
If ${\bf p}$ and $E({\bf p})$ are the  momentum and energy
of the bare particle (atom with mass $m$) measured in the
system moving with velocity
${\bf u}$,  then from the Galilean invariance it follows
that its momentum and energy measured by the observer at
rest are correspondingly
\begin{equation}
\tilde{\bf p} ={\bf p} + m {\bf u}~,~\tilde E({\bf p})
=E({\bf p}+m{\bf u})= E({\bf p}) + {\bf p}\cdot {\bf u} +
{1\over 2}m {\bf u}^2~.
\label{ParticleTransformationLaw}
\end{equation}
   This transformation law contains the mass $m$ of
the bare atom.

However,  when the quasiparticles are concerned, one can
expect that such characteristic of the microscopic  world
as the bare mass $m$ cannot enter the transformation law
for quasiparticles. This is because quasiparticles in
effective low-energy theory have no information on the
transPlanckian world of the bare atoms comprising the
vacuum state. All the information
on the quantum vacuum, which the low-energy quasiparticle
has, is encoded in the effective metric
$g_{\mu\nu}$. Since the mass $m$ must drop out from  the
transformation law for quasiparticles, we expect that the
momentum of quasiparticle is invariant under the Galilean
transformation:
${\bf p}\rightarrow {\bf p}$, while the  quasiparticle
energy is simply Doppler shifted:  $E({\bf p})\rightarrow
E({\bf p}) +{\bf p}\cdot {\bf u}$.   Such a transformation
law allows us to write the energy of quasiparticle in the
moving superfluid vacuum. If ${\bf p}$ and $E({\bf p})$
are the quasiparticle momentum and energy measured in the
coordinate system where the superfluid vacuum is at rest
(i.e.
${\bf v}_{\rm s}=0$, we call this frame the superfluid
comoving frame), then its momentum and energy in the laboratory frame are
\begin{equation}
\tilde {\bf p}={\bf p}~,~\tilde E({\bf p})=E({\bf p})+{\bf p}\cdot {\bf
v}_{\rm s}~.
\label{DopplerShiftInSuperfluids}
\end{equation}

The difference in the transformation properties of  bare
particles and quasiparticles comes from their different
status. While the momentum and energy of bare particles
are determined in ``empty'' space-time, the momentum and
energy of quasiparticles are counted from that of the
quantum vacuum. This difference can be easily visualized
if one considers the spectrum of quasiparticles in the
weakly interacting Bose gas in
Eq.(\ref{BogoliubovEnergySpectrum}) in the limit of large
momentum $p\gg mc$, when the energy spectrum of
quasiparticles approaches that of particles, $E\rightarrow
p^2/2m$. In this limit the difference between particles
and quasiparticles disappears, and at first glance one may
expect that quasiparticle should obey the same
transformation property under Galilean transformation as a
bare isolated particle. To add more confusion let us
consider an ideal Bose gas of noninteracting bare
particles, where quasiparticles have exactly the same
spectrum as particles.  Why the transformation properties
are so different for them?

The ground state of the ideal Bose gas has zero energy
and zero momentum in the reference frame where the Bose
condensate is at rest (the superfluid comoving reference
frame). In the laboratory frame the condensate  momentum
and energy are correspondingly
\begin{eqnarray}
 \left<{\cal P}\right>_{{\rm vac}}=Nm{\bf v}_{\rm s}~,
\label{GalileanParticleParticleP}\\
\left<{\cal H}\right>_{{\rm vac}}=N{m{\bf v}_{\rm s}^2\over 2} ~.
\label{GalileanParticleParticleE}
\end{eqnarray}
The state with one quasiparticle is the
state in which $N-1$ particles have zero momenta,  ${\bf
p}=0$, while one particle has nonzero momentum ${\bf
p}\neq 0$. In the comoving reference frame the momentum
and energy of such state with one quasiparticle are
correspondingly
$\left<{\cal P}\right>_{{\rm vac}+1qp}={\bf p}$ and
$\left<{\cal H}\right>_{{\rm vac}+1qp}=E({\bf p})=p^2/2m$. In the
laboratory frame
the momentum and energy of the system are obtained by Galilean transformation
\begin{eqnarray}
\left<{\cal P}\right>_{{\rm vac}+1qp}=(N-1)m{\bf
v}_{\rm s} + ({\bf p} + m{\bf v}_{\rm s})= \left<{\cal P}\right>_{{\rm vac}}+
{\bf p} ~,
\label{GalileanParticleQuasiparticleP}\\
\left<{\cal H}\right>_{{\rm vac}+1qp}=(N-1){m{\bf v}_{\rm s}^2\over 2} +{({\bf
p} + m{\bf v}_{\rm s})^2\over 2m}=\left<{\cal H}\right>_{{\rm vac}}+ E({\bf
p})+{\bf p}\cdot  {\bf v}_{\rm s}~.
\label{GalileanParticleQuasiparticleE}
\end{eqnarray}
Since the energy and the momentum of
quasiparticles are counted from that of the quantum
vacuum, the transformation properties of quasiparticles
are different from the Galilean transformation law. The
part of the Galilean transformation, which contains the
mass of the atom, is absorbed by the Bose-condensate which
represents the quantum vacuum.

\subsection{Effective metric from Galilean transformation}
\label{EffectiveMetricGalileanTransformation}

The right hand sides  of
Eqs.(\ref{GalileanParticleQuasiparticleP}) and
(\ref{GalileanParticleQuasiparticleE}) show that the
energy spectrum of quasiparticle in the moving
superfluid vacuum is given by
Eq.(\ref{DopplerShiftInSuperfluids}). Such spectrum can be
written in terms of the effective acoustic metric:
\begin{equation}
(\tilde E-{\bf p}\cdot {\bf v}_{\rm
s})^2=c^2p^2~~, ~~{\rm or}~~ g^{\mu\nu}p_\mu p_\nu=0~.
\label{PhononEnergySpectrumConformal}
\end{equation}
where the metric has the form:
\begin{eqnarray}
 g^{00}=-1 ~,~ g^{0i}=- v_{\rm s}^i ~,~
g^{ij}=  c^2\delta^{ij} -v_{\rm s}^i v_{\rm s}^j  ~,
\label{ContravarianAcousticMetric}\\
 g_{00}=- \left(1-{{\bf v}_{\rm s}^2\over c^2}\right) ~,~
g_{0i}=- {v_{{\rm s} i}\over c^2} ~,~ g_{ij}= {1\over
c^2}\delta_{ij} ~,\label{CovarianAcousticMetric}\\
\sqrt{-g}=c^{-3}~.
\label{DeterminantAcousticMetric}
\end{eqnarray}
The Eq.(\ref{PhononEnergySpectrumConformal}) does not
determine the conformal factor. The derivation of the
acoustic metric with the correct conformal factor can be
found in
Refs.\cite{Vissersonic,StoneIordanskii,StoneMomentum}.

\subsection{Broken Galilean invariance}
\label{Broken Galilean invariance}

The modified transformation  law for quasiparticles is the
consequence of the fact that the mere presence of the gas
or liquid with nonzero number $N$ of atoms breaks the
Galilean invariance. While for the total system, quantum
vacuum  + quasiparticles, the Galilean invariance is a
true symmetry, it is not applicable to the
subsystem of quasiparticles if it is considered
independently on the quantuam vacuum. This is the general
feature of the broken symmetry: the vacuum breaks the
Galilean invariance. This means that in the Bose gas and
in the superfluid
$^4$He, two symmetries are broken: the global
$U(1)$ symmetry and the Galilean invariance.
\subsection{Momentum vs pseudomomentum}
\label{MomentumPseudomomentum}

On the other hand, due to the presence of quantum vacuum,
there are two different types of translational invariance
at $T=0$ (see detailed discussion in Ref.
\cite{StoneMomentum}): (i) Invariance under the
translation of the quantum vacuum with respect to the empty
space; (ii) Invariance under translation of quasiparticle
with respect to the quantum vacuum.

The operation (i) leaves the action invariant provided
that the empty space is homogeneous. The
conserved quantity, which comes from the translational
invariance with respect to the empty space is the momentum.
The operation (ii) is the symmetry operation if the quantum
vacuum is homogeneous. This symmetry gives rise to the
pseudomomentum. Accordingly the bare particles in empty
space are characterized by the momentum, while
quasiparticles -- excitations of the quantum vacuum -- are
characterized by pseudomomentum. That is why the different
transformation properties for momentum of
particles in Eq.(\ref{ParticleTransformationLaw}) and
quasiparticles in Eq.(\ref{DopplerShiftInSuperfluids}).

The Galilean invariance is the symmetry of the underlying
microscopic physics of atoms in empty space.  It is broken
and fails to work for quasiparticles. Instead, it produces
the transformation law in
Eq.(\ref{DopplerShiftInSuperfluids}), in which the
microscopic quantity -- the mass $m$ of bare particles --
drops out. This is an example of how the memory on the
microscopic physics is erased in the low-energy corner.
Furthemore, when the low-energy corner is approached and
the effective field theory emerges, these modified
transformations gradually become the part of the more
general coordinate transformations appropriate for the
Einstein theory of gravity.

\subsection{Vacuum energy of weakly interacting Bose gas}

The vacuum energy of the Bose gas as a function of the
chemical potential
$\mu$ is
\begin{equation}
\left<{\cal H}-\mu {\cal N}\right>_{\rm vac}= -{\mu^2
\over 2U}V+ {1\over 2}\sum_{{\bf p}}\left(E({\bf p}) -
{p^2\over 2m}-mc^2 +{m^3c^4\over p^2}\right)
\label{VacuumEnergyBoseGas}
\end{equation}
The last term in round brackest is added to take into
account  the perturbative correction to the matrix element
$U$
\cite{Khalatnikov}. If the total number of particles is
fixed,  the corresponding vacuum energy is the function of
$N$:
\begin{eqnarray}
\left<{\cal H}\right>_{\rm vac}=E_{\rm
vac}(N)={1\over 2}Nmc^2+
\label{VacuumEnergyBoseGas2}\\
 {1\over
2}\sum_{{\bf p}}\left(E({\bf p}) -  {p^2\over 2m}-mc^2
+{m^3c^4\over p^2}\right)
\label{VacuumEnergyBoseGas1}
\end{eqnarray}

Inspection of the vacuum energy shows that it does contain
the zero point energy of the phonon field, ${1\over
2}\sum_{{\bf p}}E({\bf p})$. This divergent term is
balanced by three counterterms in
Eq.(\ref{VacuumEnergyBoseGas1}). They come from the
microscopic physics (they explicitly contain the
microscopic parameter -- the mass $m$ of atom).  This
regularization, which naturally arises in the microscopic
physics, is absolutely unclear within the effective
theory.  After the regularization, the contribution of the
zero point energy of the phonon field in
Eq.(\ref{VacuumEnergyBoseGas1}) becomes
\begin{equation}
{1\over
2}\sum_{{\bf p}~{\rm reg}}E({\bf p})={1\over
2}\sum_{{\bf p}}E({\bf p}) - {1\over
2}\sum_{{\bf p}}\left( {p^2\over 2m}+mc^2 -{m^3c^4\over
p^2}\right)={8\over 15\pi^2} Nmc^2 {m^3c^3\over n}~,
\label{ZerPointContribution}
\end{equation}
  where $n=N/V$ is
particle density in the vacuum. Thus the total vacuum
energy
\begin{eqnarray}
E_{\rm vac}(N)\equiv  \epsilon (n)~V=
\label{VacuumEnergyBoseGas3}\\
{1\over 2}Vmc^2
\left(n+{16\over 15\pi^2}   {m^3c^3\over \hbar^3} \right)=
\label{VacuumEnergyBoseGas4}\\
V\left({1\over 2}Un^2+{8\over 15\pi^2 \hbar^3} m^{3/2}
U^{5/2} n^{5/2}\right)
\label{VacuumEnergyBoseGas5}
\end{eqnarray}
In the weakly interacting Bose gas the contribution of the
phonon zero point motion (the second terms in
Eqs.(\ref{VacuumEnergyBoseGas4}) and
(\ref{VacuumEnergyBoseGas5})) is much smaller than the
leading contribution to the vacuum energy, which comes
from interaction (the first terms in
Eqs.(\ref{VacuumEnergyBoseGas4}) and
(\ref{VacuumEnergyBoseGas5})). The small parameter, which regulates the
perturbation theory in the above procedure is
$mca/\hbar \ll 1$ (where $a$ is the interatomic distance: $a\sim n^{-1/3}$), or
$mU/\hbar^2a\ll 1$. Small speed of sound reflects the
smallness of the pair interaction $U$.

\subsection{Planck energy scales}

The microscopic physics also shows that there are two energy parameters, which
play the role of the Planck energy scale:
\begin{equation}
E_{\rm Planck~1}=mc^2~~,~~E_{\rm
Planck~2}={\hbar c\over a}~.
\label{TwoPlanckScales}
\end{equation}
The Planck mass, which corresponds to the first Planck
scale $E_{\rm Planck~1}$, is the mass of Bose particles
$m$, that comprise the vacuum. The second Planck scale
$E_{\rm Planck~2}$ reflects the discreteness of the vacuum: the
microscopic parameter, which enters this scale, is the
mean distance between the particles in the vacuum. The
second energy scale corresponds to the Debye temperature in
solids. In a given case of weakly interacting particles
one has
$E_{\rm Planck~1}\ll E_{\rm
Planck~2}$, i.e. the distance between the particles in the vacuum is so
small, that the quantum effects are stronger than interaction. This is the
limit of strong correlations and weak interactions.

Below the first
Planck scale
$E\ll E_{\rm Planck~1}=mc^2$, the energy spectrum of quasiparticles is linear,
which corresponds to the relativistic field theory arising in the low-energy
corner. At this Planck scale the ``Lorentz'' symmetry is violated.
The first Planck scale $E_{\rm Planck~1}=mc^2$ also determines the
convergence of the sum in Eq.(\ref{VacuumEnergyBoseGas1}).
In terms of this scale the
Eq.(\ref{VacuumEnergyBoseGas1}) can be written as
\begin{equation}
V {8\over 15\pi^2}\sqrt{-g}E_{\rm Planck~1}^4 ~,
\label{CosmologicalTermBoseGas1}
\end{equation}
where $g=-1/c^6$  is the determinant of acoustic metric in
Eq.(\ref{DeterminantAcousticMetric}). This contribution to
the vacuum energy has the same structure as the
cosmological term in Eq.(\ref{VacuumEnergyPlanck2}).
However, the leading term in the vacuum energy,
Eq.(\ref{VacuumEnergyBoseGas2}), is higher and is
determined by both Planck scales:
\begin{equation}
  {1\over 2}V\sqrt{-g}E_{\rm Planck~2}^3E_{\rm Planck~1} ~.
\label{CosmologicalTermBoseGas2}
\end{equation}

\subsection{Vacuum pressure and
cosmological constant}

The relevant vacuum energy of the grand ensemble
of particles is the thermodynamic potential at fixed chemical
potential: $\left<{\cal H}-\mu {\cal N}\right>_{\rm vac}$.   It is related
to the
pressure of the liquid as (see the prove of this
thermodynamic equation  below, Eq.(\ref{VacuumEnergyPressure}))
\begin{equation}
P= -{1\over V} \left<{\cal H}-\mu {\cal N}\right>_{\rm vac} ~.
\label{ThermpPotVsPressure}
\end{equation}
Such relation between pressure and energy is similar to that in
Eq.(\ref{EquationOfState}) for the equation of state of the relativistic
quantum
vacuum, which is described by the cosmological constant.

This vacuum energy for the weakly interacting Bose gas is
given by
\begin{equation}
\left<{\cal H}-\mu {\cal N}\right>_{\rm vac}= {1\over
2}V\sqrt{-g}\left(-E_{\rm Planck~2}^3E_{\rm Planck~1}
+{16\over 15\pi^2}E_{\rm Planck~1}^4\right)~.
\label{ThermpPotEnergyBoseGas}
\end{equation}
Two terms in Eq.(\ref{ThermpPotEnergyBoseGas})  represent
two contributions to the vacuum pressure in the weakly
interacting Bose gas.  The zero point energy of the phonon
field, the second term in
Eq.(\ref{ThermpPotEnergyBoseGas}), which coincides with
Eq.(\ref{ZerPointContribution}), does lead to the negative
vacuum pressure as is expected from the effective theory.
However, the  magnitude of this negative   pressure is
smaller than the positive pressure coming from the
microscopic ``trans-Planckian'' degrees of freedom (the
first term in Eq.(\ref{ThermpPotEnergyBoseGas}) which
is provided by the repulsive interaction of atoms). Thus
the weakly interacting Bose-gas can exist only under
positive external pressure.

\section{Quantum liquid}\label{Quantum liquid}

\subsection{Real liquid $^4$He}\label{RealLiquidHelium}

In the real liquid $^4$He the  interaction between the
particles (atoms) is not small. It is strongly correlated
and strongly interacting system, where the two Planck
scales are of the same order, $mc^2\sim \hbar c/a$. This
means that the interaction energy and the energy of
zero-point motion of atoms are of the same order. This
is not the coincidince but reflects the stability og
the liquid state. Each of the two energies depend on the
particle density $n$. One can find the value of $n$ at
which the two contributions to the vacuum pressure
compensate each other. This means that the system can be in
equilibrium even at zero external pressure,
$P=0$, i.e. the quantum liquid can exist as a completely
autonomous isolated system without any interaction with
environment. This is what we must expect from the quantum
vacuum in cosmology, since there are no exteranl
environment for the vacuum.

In case of the collection of big but finite number $N$ of
$^4$He atoms at $T=0$, they do not fly away as it happens
for gases, but are held together to form a droplet of
liquid with a finite mean particle density $n$. This
density $n$ is fixed by the attractive interatomic
interaction and repulsive zero point oscillations of
atoms, only a part of this zero point motion being
described in terms of the zero point energy of phonon
mode.

The only macroscopic quantity which characterizes the homogeneous stationary
liquid at
$T=0$ is the mean particle density $n$. The vacuum energy density is the
function of
$n$
\begin{equation}
\epsilon(n)={1\over V}\left<{\cal H}\right>_{\rm vac}~,
\label{EnergyOnDensity}
\end{equation}
and this function determines the equation of state  of the
liquid. The relevant vacuum energy density -- the density
of the thermodynamic potential of grand ensemble
\begin{equation}
\tilde\epsilon(n)=\epsilon(n) -\mu n={1\over V}\left<{\cal H}-\mu {\cal
N}\right>_{\rm vac}~.
\label{ThermodynamicPotentialOnDensity}
\end{equation}
Since the particle number $N=nV$ is conserved,
$\tilde\epsilon(n)$ is the right quantity which must be
minimized to obtain the equilibrium state of the liquid at
$T=0$ (the equilibrium vacuum). The chemical potential
$\mu$ plays the role of the Lagrange multiplier
responsible for the conservation of bare atoms. Thus an
equilibrium number of particles $n_0(\mu)$ is obtained from
equation:
\begin{equation}
{d\tilde\epsilon\over dn}=0~,~{\rm or} ~~{d\epsilon\over dn}=\mu~.
\label{EquationForEquilibriumDensity}
\end{equation}
Here we discuss only spatially homogeneous ground state,
i.e. with spatially homogeneous $n$, since we know that
the ground state of helium at $T=0$ is homogeneous: it is
uniform liquid, not the crystal.

From the definition of the pressure,
\begin{equation}
P=-{d(V\epsilon(N/V))\over dV}=-\epsilon(n)+n{d\epsilon\over dn}~,
\label{DefinitionOfPressure}
\end{equation}
and from Eq.(\ref{EquationForEquilibriumDensity}) for the
density $n$ in equilibrium vacuum  one obtains that in
equilibrium the vacuum energy density
$\tilde\epsilon$ and the vacuum pressure $P$ are related by
\begin{equation}
 \tilde \epsilon_{\rm vac~eq} = -P_{\rm vac}~.
\label{VacuumEnergyPressure}
\end{equation}
The thermodynamic relation between the energy and pressure in the ground
state of the quantum liquid' $P=-\tilde \epsilon$, is the same as obtained for
vacuum energy and pressure from the Einstein cosmological term. This is
because the cosmological term also does not contain derivatives.

Close to the equilibrium state one can expand the vacuum energy in terms of
deviations of particle density from its equilibrium value. Since the linear
term disappears due to the stability of the superfluid vacuum, one has
\begin{equation}
\tilde\epsilon(n)\equiv \epsilon(n)-\mu n=-P_{\rm vac} +{1\over 2} {m
c^2\over n_0(\mu)}
(n-n_0(\mu))^2~.
\label{VacuumCloseToEquil}
\end{equation}

\subsection{Gas-like vs liquid-like vacuum}\label{GasLiquidVacuum}

It is important that the vacuum of real $^4$He is not a gas-like but
liquid-like, i.e. it can be in equilibrium at $T=0$ without interaction with
the environment. Such property of the collection of atoms at $T=0$ is
determined by
the sign of the  chemical potential, if it is counted from the energy of an
isolated
$^4$He atom.
$\mu$ is positive in a weakly interacting Bose gas, but is negative in a
real $^4$He
where $\mu\sim -7$ K \cite{Woo}.

Due to the negative $\mu$ the isolated atoms are
collected together forming the liquid droplet which is self sustained
without any
interaction with the outside world. If the droplet is big enough, so that
the surface
tension can be neglected compared to the volume effects,  the pressure in
the liquid
is absent,
$P_{\rm vac}=0$, and thus the vacuum energy density $\tilde\epsilon$
is zero in equilibrium:
\begin{equation}
\tilde\epsilon_{\rm vacuum~of~self-sustaining~system} \equiv 0~.
\label{ZeroVacuumEnergy}
\end{equation}
This condition cannot be fulfilled for gas-like states for
which
$\mu$ is positive and thus they cannot exist without an external pressure.

\subsection[Model liquid state]{Model liquid state}

It is instructive to discuss some model energy density
$\epsilon(n)$ describing a stable isolated liquid at
$T=0$. Such a model must satisfy the following condition:
(i) $\epsilon(n)$ must be attractive (negative) at small
$n$ and repulsive (positive) at large $n$ to provide
equilibrium density of liquid at intermediate $n$; (ii) The
chemical potential must be negative to prevent
evaporation; (iii) The liquid must be locally stable, i.e.
the eigen frequencies of collective modes must be real.

All these conditions can be satisfied if we modify the
Eq.(\ref{VacuumEnergyBoseGas5}) in the following way. Let
us change the sign of the first term describing
interaction and leave the second term coming from vacuum
fluctuations intact assuming that it is valid even at high
density of particles.  Due to the attractive interaction
at low density the Bose gas collapses forming the liquid
state.  Of course, this is rather artifical construction,
but it qualitatively desribes the liquid state. So we come
to the following model
\begin{equation}
\epsilon(n)= -{1\over 2}\alpha n^2+{2\over 5}\beta n^{5/2}~,
\label{ModelEnergy}
\end{equation}
though, in addition to $\alpha$ and $\beta$,  one can use
also the exponents of
$n$ as the fitting parameter. An equilibrium particle
density in terms of chemical potential is obtained from the
minimization of the relevant vacuum energy
$\tilde\epsilon=\epsilon -\mu n$ over $n$:
\begin{equation}
{d\epsilon\over dn}=\mu ~\rightarrow~~  -\alpha n_0 +\beta n_0^{3/2}=\mu
\label{ModelEquilibrium}
\end{equation}
The equation of state of such a liquid is
\begin{equation}
P(n_0)=-\left(\epsilon(n_0) -\mu n_0\right)= -{1\over
2}\alpha n_0^2+{3\over 5}\beta n_0^{5/2}
\label{PressureVsDensity}
\end{equation}
This equation of state allows the existence of the
isolated liquid droplet, for which an external pressure is
zero, $P=0$. The equilibrium density, chemical potential
and speed of sound in the isolated liquid are
\begin{eqnarray}
n_0(P=0)= \left({5\alpha\over 6\beta}\right)^2 ~,
\label{EquilibriumDensityAtZeroP}\\
\mu(P=0)=-{1\over 6} n_0 \alpha ~,
\label{EquilibriumChemicalPotentialAtZeroP}\\
~mc^2=\left({dP\over
dn_0}\right)_{P=0}=\left(n{d^2 \epsilon\over
dn^2}\right)_{P=0}={7\over 8} n_0 \alpha =5.25 ~|\mu|~.
\label{EquilibriumSpeedSoundAtZeroP}
\end{eqnarray}
This liquid state is stable: the chemical potential $\mu$ is negative
preventing
evaporation, while
$c^2$ is positive, i.e. the compressibility is negative, which indicates
the local
stability of the liquid.

The Eq.(\ref{PressureVsDensity}) shows that the quantum
zero point energy produces a positive contribution to the
vacuum pressure, instead of the negative pressure expected
from the effective theory and from
Eq.(\ref{ThermpPotEnergyBoseGas}) for the weakly
interacting Bose gas. Let us now recall that in this model
we changed the sign of the interaction term, compared to
that in the weakly interacting Bose gas. As a result both
terms in Eq.(\ref{ThermpPotEnergyBoseGas}) have changed
sign.

The equilibrium state of the liquid is obtained due to the
competition of two effects: attractive interaction of bare
atoms (corresponding to the negative vacuum pressure in
Eq.(\ref{PressureVsDensity})) and their zero point motion
which leads to repulsion  (corresponding  to the positive
vacuum pressure in Eq.(\ref{PressureVsDensity})). These
effects are balanced in equilibrium, that is why the two
``Planck'' scales in Eq.(\ref{TwoPlanckScales}) become of
the same order of magnitude.

\subsection{Quantum liquid from Theory of Everything}

The parameters of liquid $^4$He at $P=0$ have been calculated in exact
microscopic
theory, where the many-body wave function of $^4$He atoms has been constructed
using the ``Theory of Everything'' in Eq.(\ref{TheoryOfEverything}) with
realistic
pair potential  \cite{Woo}. For $P=0$ one has
\begin{eqnarray}
n_0\sim 2\cdot 10^{22}~{\rm
cm}^{-3}~,~\mu={\epsilon(n_0)\over n_0}
\sim -7~{\rm K}~,~c\sim 2.5\cdot 10^4~{\rm cm}/{\rm sec}~,
\label{ParametersMicroscopicTheory}\\
  mc^2\sim
30~{\rm K}~, ~ \hbar cn_0^{1/3}\sim 7~{\rm K}
~,\label{PlanckScaleExactTheory}\\
~\tilde\epsilon
\equiv 0~.
\label{AgainVacuumEnergy}
\end{eqnarray}
These derived parameters are in a good agreement with
their experimental values.

\section{Vacuum energy and cosmological constant}

\subsection{Nullification
of ``cosmological constant''  in quantum liquid}

If there is no interaction with environment, the external
pressure $P$ is zero, and thus in
equilibrium the vacuum energy density $\epsilon
-\mu n=-P$ in Eqs.(\ref{ThermpPotVsPressure}) and
(\ref{VacuumEnergyPressure}) is also zero. The energy
density $\tilde\epsilon$ is the quantity which is relevant
for the effective theory: just this energy density enters
the effective action for the soft variables, including the
effective gravity field, which must be minimized to obtain
the stationary states of the vacuum and matter fields. Thus
$\tilde\epsilon$ is the proper counterpart of the vacuum
energy density, which is responsible for the cosmological
term in the Einstein gravity.

Nullification of both the vacuum energy density and the
pressure in the quantum liquid means that
$P_{\Lambda}=-\rho_{\Lambda}=0$, i.e. the effective
cosmological constant in the liquid is identically zero.
Such nullification of the cosmological constant occurs
without any fine-tuning or supersymmetry. Note that the
supersymmetry -- the symmetry between the fermions and
bosons -- is simply impossible in
$^4$He, since there are no fermionic fields
in the Bose liquid.  The same
nullification occurs in Fermi liquids, in superfluid phases
of $^3$He, since these are also the quantum liquids with
the negative chemical potential \cite{PhysRepRev}. Some
elements of supersymmetry can be found in the effective
theory of superfluid $^3$He
\cite{Nambu,PhysRepRev}, but this is certainly not enough
to produce the nullification.

Applying this to the quantum vacuum, the mere assumption
that the ``cosmological liquid'' -- the vacuum of the
quantum field theory -- belongs to the class of states,
which can exist in equilibrium without external forces,
leads to the nullification of the vacuum energy in
equilibrium at $T=0$.

Whether this scenario of nullification of cosmological
constant can be applied to the cosmological fluid (the
physical vacuum) is a question under discussion (see
discussion in Ref. \cite{WhoIsInflaton}, where the
inflaton field is considered as the analog of the variable
$n$ in quantum liquid).

\subsection{Role of zero point energy of bosonic and
fermionic fields}

The advantage of the quantum liquid is that we
know both the effective theory and the fundamental
Theory of Everything in Eq.~(\ref{TheoryOfEverything}).
That is why we can compare the two approaches.
The microscopic wave function used for microscopic
calculations contains, in principle, all the information on
the system, including the quantum fluctuations of the
low-energy phonon degrees of freedom, which are considered
in the effective theory in
Eq.(\ref{VacuumEnergyEffective}).   That is why the
separate treatment of the contribution to the vacuum
energy of the low-energy degrees of freedom described by
effective theory has no sense: this leads at best to the
double counting.

The effective theory in quantum Bose liquid contains
phonons as elementary bosonic quasiparticles and no
fermions. That is why the analogue of
Eq.~(\ref{VacuumEnergyPlanck}) for the vacuum energy
produced by the zero point motion of ``elementary
particles'' is
\begin{equation}
\rho_\Lambda = {1\over 2V}\sum_{\rm phonons}cp \sim
{1\over c^3} E^4_{\rm Planck}=\sqrt{-g}~E^4_{\rm Planck}~.
\label{VacuumEnergyEffective}
\end{equation}
Here $g$ is the determinant of the acoustic
metric in Eq.~(\ref{DeterminantAcousticMetric}).
The ``Planck'' energy cut-off can be chosen either as the
Debye temperature $E_{\rm Debye}=\hbar c/a=\hbar
cn_0^{1/3}$ in Eq.(\ref{PlanckScaleExactTheory})  with
$a$ being the interatomic distance, which plays the role of
the Planck length; or as $mc^2$ which has the same order of
magnitude.

The disadvantages of such a naive calculation of the vacuum
energy within the effective field theory  are: (i) The
result depends on the cut-off procedure; (ii) The result
depends on the choice of the zero from which the energy is
counted: a shift of the zero level leads to a shift in the
vacuum energy.

In the microscopic theory these disadvantages are cured:
(i) The cut-off is not required; (ii) The relevant  energy
density,
$\tilde\epsilon =\epsilon - \mu n$, does not depend on the
choice of zero level: the shift of the energy
$\int d^3r \epsilon$ is exactly compensated by the shift of
the chemical potential $\mu$.

At $P=0$ the microscopic results for both vacuum energies
characterizing the quantum liquid are:
$\tilde\epsilon(n_0)=0$, $\epsilon(n_0)=\mu n_0 <0$. Both
energies are in severe disagreement with the naive
estimation in Eq.(\ref{VacuumEnergyEffective}) obtained
within the effective theory:  $\rho_\Lambda$ in
Eq.(\ref{VacuumEnergyEffective}) is nonzero in
contradiction with $\tilde\epsilon(n_0)=0$; comparing it
with $\epsilon(n_0)$ one finds that $\rho_\Lambda$ is about
of the same order of magnitude, but it has an opposite
sign.

This is an important lesson from the
condensed matter. It shows that the use of the zero point
fluctuations of bosonic or fermionic modes in
Eq.(\ref{VacuumEnergyPlanck})  in the cis-Planckian
effective theory is absolutely irrelevant for the
calculations of the vacuum energy density. Whatever
are the low-energy modes, fermionic or bosonic, for
equilibrium vacuum they are exactly cancelled by the
transn-Planckian degrees of freedom, which are not
accessible  within the effective theory.

\subsection{Why is equilibrium vacuum not gravitating?}

We discussed the condensed matter view to the problem,
why the vacuum energy is so small, and found that the
answer comes from the ``fundamental trans-Planckian
physics''. In the effective theory of the low energy
degrees of freedom the vacuum energy density of a quantum
liquid is of order $E_{\rm Planck}^4$ with the
corresponding ``Planck'' energy appropriate for this
effective theory. However, from the exact ``Theory of
Everything'' of the quantum liquid, i.e. from the
microscopic physics, it follows that the
``trans-Planckian'' degrees of freedom exactly cancel the
relevant vacuum energy without fine tuning. The vacuum
energy density  is exactly zero, if the following
conditions are fulfilled: (i) there are no external forces
acting on the liquid; (ii) there are no quasiparticles
(matter) in the liquid;  (iii) no curvature or
inhomogeneity in the liquid; and (iv) no boundaries which
give rise to the Casimir effect. Each of these factors
perturbs the vacuum state and induces a nonzero value of
the vacuum energy density of order of the energy density
of the perturbation, as we shall discuss below.

Let us, however, mention, that the actual problem for
cosmology is not why the vacuum energy is zero (or very
small when it is perturbed), but why the vacuum is not (or
almost not) gravitating.  These two problems are not
necessarily related since in the effective theory the
equivalence principle is not the fundamental physical law,
and thus does not necessarily hold when applied to the
vacuum energy. That is why, one cannot exclude the
situation, when the vacuum energy is huge, but it is not
gravitating. The condensed matter provides an example of
such situation too. The weakly interacting Bose gas
discussed above is just the proper object. This
gas-like substance can exists only at positive external
pressure, and thus it has the negative energy density. The
translation to the relativistic language gives a huge
vacuum energy is on the order of the Planck energy scale
(see Eq.(\ref{ThermpPotEnergyBoseGas})). Nevertheless,
the effective theory remains the same as for the
quantum liquid, and thus even in this situation the
equilibrium vacuum, which exists under an external
pressure, is not gravitating, and only small deviations
from equilibrium state are gravitating. Just this
situation was discussed in Ref. \cite{WhoIsInflaton}.

In condensed matter the effective gravity appears as an
emergent phenomenon in the low energy corner. The
gravitational field is not fundamental, but is one of the
low energy collective modes of the quantum vacuum.  This
dynamical mode provides the effective metric (the acoustic
metric in $^4$He and weakly interacting Bose gas) for the
low-energy quasiparticles which serve as an analogue of
matter. This gravity does not exist on the microscopic
(trans-Planckian) level and appears only in the low energy
limit together with the ``relativistic'' quasiparticles
and the acoustics itself. The bare atoms, which live in the
``trans-Planckian'' world and form the vacuum state
there,  do not experience the ``gravitational'' attraction
experienced by the low-energy quasiparticles, since the
effective gravity simply does not exist at the micriscopic
scale (we neglect here the real gravitational attraction
of the atoms, which is extremely small in quantum
liquids). That is why the vacuum energy cannot serve as a
source of the effective gravity field: the pure completely
equilibrium homogeneous vacuum is not gravitating.

On the other hand, the long-wave-length perturbations of
the vacuum are within the sphere of influence of the
low-energy effective theory; such perturbations can be the
source of the effective gravitational field. Deviations of
the vacuum from its equilibrium state, induced by
different sources discussed below, are gravitating.

\subsection{Why is the vacuum energy unaffected by the
phase transition?}

It is commonly  believed that the vacuum of the Universe
underwent one or several broken symmetry phase
transitions. Since each of the transtions is accompanied
by a substantial change in the vacuum energy, it is not
clear why the vacuum energy is (almost) zero after the
last phase transition. In other words, why has the true
vacuum the zero energy, while the energies of all other
false vacua are enormously big?

What happens in quantum liquids? According to the
conventional wisdom, the phase transition, say, to the
broken symmetry vacuum state, is accompanied by the change
of the vacuum energy, which must decrease in a phase
transition. This is what usually follows from the
Ginzburg-Landau description of phase transitions.
However, let us compare the energy densities of
the false and the true vacuum states. Let us assume that
the phase transition is of the first order, and the false
vacuum is separated from the true vacuum by a large energy
barrier, so that it can exist as a (meta)stable state.
Since the false vacuum is stable, the
Eq.~(\ref{ZeroVacuumEnergy}) can also be applied to the
false vacuum, and one obtains the paradoxical result: in
the absence of external forces the energy density of the
false vacuum must be the same as the
energy density of the true vacuum, i.e. the
relevant energy density $\tilde\epsilon$ must be zero for
both vacua. Thus the first order phase transition occurs
without the change in the vacuum energy.

To add more confusion, note that the
Eq.~(\ref{ZeroVacuumEnergy}) can be applied even
to the unstable vacuum which corresponds to a saddle point
of the energy functional, if such a vacuum state can live
long enough.  Thus the vacuum energy density does not
change in the second order phase transition either.

There is no paradox, however: after the phase transition to
a new state has occured, the chemical potential $\mu$ will
be automatically ajusted to preserve the zero external
pressure and thus the zero energy $\tilde\epsilon$ of the
vacuum. Thus the relevant vacuum energy is zero before and
after transition, which means that the $T=0$ phase
transitions do not disturb the zero value of the
cosmological constant. Thus the scenario of the
nullification of the vacuum energy suggested by the
quantum liquids survives even if the phase transition
occurs in the vacuum. The first order phase transition
between superfluid phases $^3$He-A and $^3$He-B at $T=0$
and $P=0$ gives the proper example \cite{PhysRepRev}.

\subsection{Why is the cosmological constant nonzero?}

We now come to another problem in cosmology: Why is the
vacuum energy density presently of the same order of
magnitude as the energy density of matter $\rho_M$, as is
indicated by recent astronomical observations
\cite{Supernovae}. While the relation between
$\rho_M$ and $\rho_\Lambda$ seems to depend on the details
of trans-Planckian physics, the order of magnitude
estimation can be readily obtained. In equilibrium and
without matter the vacuum energy is zero. However, the
perturbations of the vacuum caused by matter and/or by the
inhomogeneity of the metric tensor lead to disbalance. As
a result the deviations of the vacuum energy from zero
must be on the of order of the perturbations.

Let us consider how this happens in quantum liquids for
different types of perturbations, i.e. how the vacuum
energy, which is zero at $T=0$ and in complete equilibrium
in the absence of external forces, is influenced by
different factors, which lead to small but nonzero value
of the cosmological constant.

\subsection{Vacuum energy from finite temperature}

A typical example derived  from quantum liquids is the
vacuum energy produced by temperature.  Let us consider
for example the superfluid $^4$He in equilibrium at finite
$T$ without external forces. If $T\ll -\mu$ one can
neglect the exponentially small evaporation and consider
the liquid as in equilibrium. The quasiparticles -- phonons
-- play the role of the hot relativistic matter, and their
equaton of state is
$P_{M}=(1/3)\rho_M=(\pi^2/30\hbar^3 c^3)  T^4 $,
with $c$ being the speed of sound \cite{Khalatnikov}. In
equilibrium the pressure caused by thermal quasiparticles
must be compensated by the negative vacuum pressure,
$P_{\Lambda}=-P_{M}$, to support the zero value of the
external pressure,
$P=P_{\Lambda}+P_{M}=0$.  In this case one has the
following nonzero values of the
vacuum pressure and vacuum energy density:
\begin{equation}
\rho_{\Lambda}=-P_{\Lambda}=P_{M}={1\over
3}\rho_{M}=\sqrt{-g}\frac{\pi^2}{30\hbar^3}  T^4~,
\label{VacuumMatterEnergy}
\end{equation}
where $g=-c^{-6}$  is again the determinant of acoustic
metric. In this example
the vacuum energy density
$\rho_{\Lambda}$ is positive and always on the order of the energy density of
matter. This indicates that the cosmological constant is not actually a
constant
but is ajusted to the energy density of matter and/or to the other
perturbations
of the vacuum discussed below.

\subsection{Vacuum energy from Casimir effect}

Another example of the induced nonzero vacuum energy density is provided by the
boundaries of the system. Let us consider a finite droplet of $^4$He with
radius $R$. If this droplet is freely suspended then at $T=0$ the vacuum
pressure
$P_{\Lambda}$ must compensate the pressure caused by the surface tension
due to the
curvature of the surface. For a spherical droplet one obtains the negative
vacuum
energy density:
\begin{equation}
\rho_{\Lambda}=-P_{\Lambda}=-{2\sigma\over
R} \sim - {E^3_{\rm Debye}\over\hbar^2c^2 R}\equiv -\sqrt{-g}E^3_{\rm
Planck}{\hbar c\over R}~,
\label{VacuumDropletEnergy}
\end{equation}
where $\sigma$ is the surface tension.  This is an
analogue of the Casimir effect, in which the boundaries of
the system produce a nonzero vacuum pressure. The strong
cubic dependence of the vacuum pressure on the ``Planck''
energy $E_{\rm Planck}\equiv E_{\rm Debye}$ reflects the
trans-Planckian origin of the surface tension $\sigma \sim
E_{\rm Debye}/a^2\sim \hbar c/a^3$: it is the energy (per
unit area) related to the distortion of atoms in the
surface layer of the size of the interatomic distance $a$.

Such term of order $E^3_{\rm Planck}/R$ in the Casimir
energy has been considered in  Ref.\cite{Ravndal}. In Ref.
\cite{Bjorken} such vacuum energy, with $R$ being the
size of the horizon, has been connected to the
energy of the Higgs condensate in the electroweak phase
transition.

This form  of the Casimir energy -- the surface energy
$4\pi R^2\sigma$ normalized to the volume of the droplet
-- can also serve as an analogue of the quintessence in
cosmology
\cite{Quintessence}. Its equation of state is
$P_{\sigma} =  -(2/3)\rho_{\sigma}$:
\begin{equation}
\rho_{\sigma}={4\pi R^2 \sigma \over {4\over 3} \pi
R^3}={3 \sigma\over R}~~,~~P_{\sigma} =-{2\sigma\over R}= -{2\over 3}
\rho_{\sigma}~.
\label{VacuumDropletEnergy2}
\end{equation}
The equilibrium condition within  the droplet can be
written as
$P=P_{\Lambda}+P_{\sigma}=0$.  In this case the
quintessence is related to the wall -- the boundary of the
droplet. In cosmology the quintessence with the same
equation of state,
$<P_{\sigma}> =  -(2/3)<\rho_{\sigma}>$,  is represented
by a wall wrapped around the Universe or by a tangled
network of cosmic domain walls \cite{TurnerWhite}. The
surface tension of the cosmic walls can be much smaller
than the Planck scale.

\subsection{Vacuum energy induced by texture}

The nonzero vacuum energy density, with a weaker dependence on $E_{\rm
Planck}$,
is induced by the inhomogeneity of the vacuum. Let us discuss the vacuum energy
density induced by texture in a quantum liquid. We consider here
the twist soliton in $^3$He-A, since such texture is
related to the Riemann curvature in general relativity
\cite{PhysRepRev}.  Within the soliton the field of the
$^3$He-A order parameter --  the unit vector
$\hat{\bf l}$ -- has a form  $\hat{\bf l}(z)=\hat{\bf
x}\cos\phi(z)+ \hat{\bf y}\sin\phi(z)$. The energy of the
system in the presence of the soliton consists of the
vacuum energy
$\rho_{\Lambda}(\phi)$ and the gradient energy:
\begin{equation}
\rho= \rho_{\Lambda}(\phi)+\rho_{\rm grad}~,~
\rho_{\Lambda}(\phi)=\rho_{\Lambda}(\phi=0)+{K\over \xi_D^2 }\sin^2\phi~,~
\rho_{\rm grad}=K(\partial_z\phi)^2~,
\label{VacuumEnergySoliton}
\end{equation}
where $\xi_D$ is the so-called dipole length
\cite{VollhardtWolfle}.  Here we denoted the energy
$\tilde\epsilon$ by $\rho$ to make the connection with
general relativity, and omitted $\sqrt{-g}$ assuming that
$c=1$.

The solitonic solution of the sine-Gordon equation, $\tan(\phi/2)=e^{z/\xi_D}$,
gives the following spatial dependence of vacuum and
gradient energies:
\begin{equation}
\rho_{\Lambda}(z)-\rho_{\Lambda}(\phi=0)=\rho_{\rm grad}(z) = {K\over \xi_D^2
\cosh^2 (z/\xi_D)}~.
\label{VacuumGradientEnergy1}
\end{equation}
Let us consider for simplicity the 1+1 case.  Then the
equilibrium state of the whole quantum liquid with the
texture can be discussed in terms of partial pressure of
the vacuum,  $P_{\Lambda}=-\rho_{\Lambda}$, and that of
the  inhomogeneity,
$P_{\rm grad}=\rho_{\rm grad}$. The latter equation of state describes the so
called stiff matter in cosmology. In equilibrium the external pressure is
zero and thus the positive pressure of the texture (stiff matter) must be
compensated by the negative pressure of the vacuum:
\begin{equation}
P=P_{\Lambda}(z)+P_{\rm grad}(z)=0~.
\label{VacuumGradientEquilibrium}
\end{equation}
This equilibrium condition produces another relation between the vacuum and the
gradient energy densities
\begin{equation}
\rho_{\Lambda}(z)=-P_{\Lambda}(z)=P_{\rm grad}(z) =\rho_{\rm
grad}(z)~.
\label{VacuumGradientEnergy2}
\end{equation}
Compariing this Eq.(\ref{VacuumGradientEnergy2}) with
Eq.~(\ref{VacuumGradientEnergy1}) one finds that in
equilibrium
\begin{equation}
\rho_{\Lambda}(\phi=0)=0~,
\label{AgainZero}
\end{equation}
i.e.. as before, the main vacuum energy density -- the
energy density of the bulk liquid far from the soliton --
is exactly zero if the isolated liquid is in equilibrium.
Within the soliton the vacuum is perturbed, and the vacuum
energy is induced being on the order of the energy of the
perturbation. In this case
$\rho_{\Lambda}(z)$ is equal to the gradient energy
density of the texture.

The induced vacuum energy density
in Eq.~(\ref{VacuumGradientEnergy1}) is inversly
proportional to the square of the size of the region where
the field is concentrated:
\begin{equation}
\rho_{\Lambda}(R)\sim
\sqrt{-g} E^2_{\rm Planck} \left({\hbar c\over
R}\right)^2~.
\label{EnergyInverseSquare}
\end{equation}
In case of
the soliton soliton $R\sim
\xi_D$. Similar behavior for the vacuum energy density in
the interior region of the Schwarzschild black hole, with
$R$ being the Schwarzschild radius, was discussed in
Ref.\cite{ChaplineLaughlin}.

In cosmology, the vacuum energy density obeying the
Eq.(\ref{EnergyInverseSquare}) with $R$ proportional to the
Robertson-Walker scale factor has been suggested in Ref.
\cite{Chapline}, and with $R$ being the size of the
horizon,
$R= R_{\rm H}$, in Ref. \cite{Bjorken}. Following the
reasoning of Ref. \cite{Bjorken}, one can state that the
vacuum energy density related to the phase transition is
determined by Eq.(\ref{EnergyInverseSquare}) with
$R= R_{\rm H}(t)$ at the cosmological
time $t$ when this transition (or crossover) occured.
Applying this to, say, the cosmological electroweak
transition, where the energy density of the Higgs
condensate is of order of
$T_{\rm ew}^4$, one obtains the relation
$T_{\rm ew}^2=E_{\rm Planck}
\hbar c/ R_{\rm H}(t=t_{\rm ew})$. It
also follows that the entropy within the horizon volume at
any given cosmological temperature
$T$ is $S_{\rm H}\sim E_{\rm Planck}^3/T^3$ for the
radiation-dominated Universe.

\subsection{Vacuum energy due to Riemann curvature}

The vacuum energy $\sim R^{-2}$, with $R$ proportional to
the Robertson-Walker scale factor, comes also from the
Riemann curvature in general relativity. It appears that
the gradient energy of a twisted $\hat{\bf l}$-texture is
equivalent to the Einstein curvature term in the action
for the effective gravitational field in $^3$He-A
\cite{PhysRepRev}:
\begin{equation}
-{1\over 16 \pi G} \int d^3r  \sqrt{-g}{\cal  R} \equiv K
 \int d^3r
((\hat{\bf l}\cdot(
\nabla\times\hat{\bf l}))^2 ~.
\label{EinsteinActionHe}
\end{equation}
Here ${\cal  R}$  is the Riemann curvature calculated
using the effective metric experienced by fermionic
quasiparticles in $^3$He-A
\begin{equation}
ds^2=-dt^2 +c_\perp^{-2}(\hat{\bf l}
\times d{\bf r})^2+c_\parallel^{-2}(\hat{\bf l}\cdot d{\bf
r})^2~.
\label{Metric3HeA}
\end{equation}
The order parameter vector $\hat{\bf l}$ plays the role of
the Kasner axis; $c_\parallel$ and $c_\perp$ correspond
to the speed of ``light'' propagating along the
direction of $\hat{\bf l}$ and in transverse direction;
$c_\parallel\gg c_\perp$.

The analogy between the textural (gradient) energy in
$^3$He-A and the curvature in general relativity allows us
to interprete the result of the previous section,
Eq.(\ref{VacuumGradientEnergy2}), in terms of the vacuum
energy induced by the curvature of the space.
It appears that in cosmology this effect can be described
within the general relativity. We must consider the
stationary cosmological model, since the time dependent
vacuum energy is certainly beyond the Einstein theory.
The stationary Universe was obtained by Einstein in his
work where he first introduced the cosmological term
\cite{EinsteinCosmCon}. It is the closed Universe with
positive curvature and with matter, where the effect of the
curvature is compensated by the cosmological term, which
is ajusted in such a way, that the Universe remains static.
This is just the correct and probably unique example, of
how the vacuum energy is induced by curvature and matter
within the general relativity.

 Let us recall this solution.  In the static
state of the Universe two equilibrium conditions must be fulfilled:
\begin{equation}
\rho=\rho_M+\rho_\Lambda+ \rho_{\cal R}=0~,~ P=P_M+P_\Lambda+ P_{\cal
R}=0~.
\label{CurvatureEnergy}
\end{equation}
The first equation in (\ref{CurvatureEnergy}) reflects the
gravitational equilibrium, which requires that the total mass density must be
zero: $\rho=\rho_M+\rho_\Lambda+ \rho_{\cal R}=0$ (actually the
``gravineutrality'' corresponds to the combination  of two equations
in (\ref{CurvatureEnergy}), $\rho+3P=0$, since $\rho+3P$ serves as a source
of the
gravitational field in the Newtonian limit). This gravineutrality is
analogous to the electroneutrality in condensed matter. The second equation in
(\ref{CurvatureEnergy}) is equivalent to the requirement that for the
``isolated''
Universe  the external pressure must be zero:
$P=P_M+P_\Lambda+ P_{\cal R}=0$.  In addition to matter density $\rho_M$ and
vacuum energy density
$\rho_{\Lambda}$, the energy density  $\rho_{\cal R}$ stored in the spatial
curvature is added:
\begin{equation}
 \rho_{\cal R}=-{{\cal R}\over 16\pi G}=-{3k\over 8\pi
GR^2}~,~ P_{\cal R} =-{1\over 3}\rho_{\cal R}~,
\label{CurvatureMass}
\end{equation}
Here $R$ is the cosmic scale factor  in the
Friedmann-Robertson-Walker metric
\begin{equation}
ds^2=-dt^2+ R^2\left({dr^2\over 1-kr^2}
+r^2d\theta^2 +r^2
\sin^2\theta d\phi^2\right)~,
\label{FriedmannRobertsonWalker}
\end{equation}
 the parameter
$k=(-1,0,+1)$  for an open, flat,  or closed Universe
respectively; and we again removed the factor $\sqrt{-g}$
from the definition of the energy densities.

For the cold Universe with $P_M=0$,  the Eqs.~(\ref{CurvatureEnergy})  give
\begin{equation}
\rho_\Lambda= {1\over 2} \rho_M =-{1\over 3}\rho_{\cal R}= {k\over 8\pi
GR^2} ~,
\label{EinsteinSolution1}
\end{equation}
and for the hot Universe with the equation of state $P_M=(1/3)\rho_M$,
\begin{equation}
\rho_\Lambda=   \rho_M =-{1\over 2}\rho_{\cal R}= {3k\over 16\pi
GR^2}~.
\label{EinsteinSolution2}
\end{equation}
Since the energy of matter is positive, the static
Universe is possible only for positive curvature,  $k=+1$,
i.e. for the closed Universe.

This is the unique solution, which describes an
equilibrium static state of the Universe, where the vacuum
energy is induced by matter and curvature. And this
solution is obtained within the effective theory of
general relativity without invoking the trans-Planckian
physics and thus does not depend on details of the
trans-Planckian physics.

\subsection{Necessity of Planck physics for
time-dependent cosmology}

The condensed matter analog  of gravity provides a natural
explanation, why the cosmological constant is zero with
a great accuracy, when compared with the result based on
naive estimation of the vacuum energy within the effective
theory. It also shows how the small effective cosmological
constant of the relative order of
$10^{-120}$ naturally arises as the response to different
perturbations. We considered the time-independent
perturbations, where the minimum energy consideration and
equilibrium condition provided the solution of the problem.

For the time-dependent situation, such as an expansion
of the Universe, the calculation of the vacuum response
is not as simple even in quantum liquids. One must solve
self-constistently the coupled dynamical equations for the
motion of the vacuum and matter fields. In case of general
relativity this requires the equation of motion for the
vacuum energy $\rho_\Lambda$, but this is certainly
beyond the effective theory since the time
dependence of $\rho_\Lambda$ violates Bianchi identities.
Probably some extension of general relativity towards the
scalar-tensor theory of gravity such as discussed in Ref.
\cite{StarobinsyPolarski}) will be more relevant for that.

On the other hand the connection to the Planck physics can
help to solve the other cosmological problems. For example
there is the flatness problem:  To arrive at the
Universe we see today, the Universe must have begun
extremely flat, which means that parameter $k$ in the
Robertson-Walker metric must be zero. In quantum
liquids the general Robertson-Walker metric in
Eq.(\ref{FriedmannRobertsonWalker}) describes the
spatially homogeneous space-time as viewed by the
low-energy quasiparticles within the effective theory.
However, for the external or high-energy observer the
quantum liquid is not homogeneous if
$k\neq 0$.   The same probably happens in gravity: If
general relativity is the effective theory, the invariance
under the coordinate transformations exists only at low
energy. For the ``Planck'' observer the Robertson-Walker
metric in Eq.(\ref{FriedmannRobertsonWalker}) is viewed as
space dependent if $k\neq 0$. That is why the condition,
that the Universe must be spatially homogeneous  not only
on the level of the effective theory but also on the
fundamental level,  requires that
$k=0$. Thus, if general relativity is the effective theory,
the truely homogeneous Universe must be flat.

\section{Effects of discrete number $N$ of particles in
the vacuum}\label{EffectsDiscreteNumberN}

\subsection{Casimir effect in quantum liquids}

Till now we used the conservation law for the particle
number $N$, the number of bare atoms in the quantum
vacuum,  to derive the nullification of the vacuum energy
in the grand ensemble of particles. Now we consider another
possible consequence of the discrete nature of the
quantim vacuum in quantum liquids. This is related to the
Casimir effect.

The attractive force between two
parallel metallic plates in vacuum induced by the quantum
fluctuations of the electromagnetic field has been
predicted by Casimir in 1948
\cite{Casimir}.  The calculation of the vacuum presure is
based on the regularization schemes, which allows to
separate the effect of the low-energy modes of the vacuum
from the huge diverging contribution of the high-energy
degrees of the freedom. There are different regularization
schemes: Riemann's zeta-function regularization;
introduction of the exponential cutoff; dimensional
regularization, etc.  People are happy when different
regularization schemes give the same results. But this is
not always so (see e.g.
\cite{Kamenshchik,Ravndal,Falomir}, and in particular
the divergencies occurring for spherical geometry in odd
spatial dimension are not
cancelled \cite{Milton,CasimirForSphere}). This raises
some criticism against the regularization methods
\cite{Hagen} or even some doubts concerning the existence
and the magnitude of the Casimir effect.

The same type of the Casimir effect arises in condensed
matter, due to thermal (see review paper \cite{Kardar})
or/and quantum fluctuations.  When considering the analog
of the Casimir effect in condensed matter, the following
correspondence must be taken into account, as we
discussed above.   The ground state of quantum liquid
corresponds to the vacuum of quantum field theory.   The
low-energy bosonic and fermionic axcitations
abobe the vacuum -- quasiparticles --
correspond to elementary particles forming the matter. The
low energy modes with linear spectrum
$E=cp$ can be described by the
relativistic-type effective theory. The analog of
the Planck energy scale
$E_{\rm Planck}$ is determined either by
the mass
$m$ of the atom of the liquid, $E_{\rm Planck}\equiv
mc^2$, or by the Debye energy, $E_{\rm Planck}\equiv
\hbar c/a$ (see Eq.(\ref{TwoPlanckScales})).

The traditional Casimir effects deals with the low energy massless modes.
The typical massless modes in quantum liquid are sound
waves. The acoustic field is desribed by the effective
theory in Eq.(\ref{LagrangianSoundWaves}) and corresponds
to the massless scalar field. The walls provide the
boundary conditions for the sound wave modes, usually these
are the Neumann boundary conditions. Because of the quantum
hydrodynamic fluctuations there must be the Casimir force
between two parallel plates immersed in the quantum
liquid. Within the effective theory the Casimir force is
given by the same equation as the Casimir force acting
between the conducting walls due to  quantum
electromagnetic fluctuations.  The only modifications are:
(i)  the speed of light must be substututed by the spin of
sound $c$;  (ii)  the factor
$1/2$ must be added, since we have the scalar field of the
longitudinal sound wave instead of  two polarizations of
light. If
$d$ is the distance between the plates and $A$ is their
area, then the
$d$-dependent contribution to the ground state energy of
the quantum liquid at $T=0$ which follows from the
effective theory must be
\begin{equation}
E_{\rm C}= -{\hbar c\pi^2 A\over 1440 d^3}
\label{CasimirForceSound}
\end{equation}
Such microscopic quantities of the quantum liquid as the
mass of the atom $m$ and interatomic space $a$
do not enter explicitly the Eq.(\ref{CasimirForceSound}):
the traditional Casimir force is completely determined by
the ``fundamental'' parameter $c$ of the effective
scalar field theory.

\subsection{Finite-size vs finite-$N$ effect}

However, we shall show that the Eq.(\ref{CasimirForceSound})
is not always true. We shall give here an example, where
the effective theory is not able to predict the Casimir
force,  since the microscopic high-energy degrees of
freedom become important. In other words the
``transPlanckian physics'' shows up and the ``Planck''
energy scale explicitly enters the result. In this situation
the Planck scale is physical and cannot be removed by
any regularization.

The Eq.(\ref{CasimirForceSound}) gives a finite-size
contribution to the energy of quantum liquid. It is
inversly proportional to the linear dimension of the
system,
$E_{\rm C} \propto 1/R$ for the sphere of radius $R$..
However, for us it is important that it is not only the
finite-size effect, but also the finite-$N$ effect, $
E_{\rm C}
\propto N^{-1/3}$, where
$N$ is the number of atoms in the liquid in the slab. As
distinct from $R$ the quantity $N$ is a discrete
quantity.  Since
the main contribution to the vacuum energy is
$\propto R^3 \propto N$, the
relative correction of order $N^{-4/3}$ means that the
Casimir force is the mesoscopic effect. We shall show that
in quantum liquids, the essentially larger mesoscopic
effects, of the relative order $N^{-1}$, can be more
pronounced.  This is a finite-$N$ effect, which
reflects the dicreteness of the vacuum and cannot be
described by the effective theory dealing with the
continuous medium, even if the theory includes the real
boundary conditions with the frequency dependence of
dielectic permeability.

We shall start with the simplest quantum vacuum  -- the
ideal one-dimensional Fermi gas -- where the mesoscopic
Casimir forces can be calculated exactly without invoking
any regularization procedure.

\subsection{Vacuum energy from microscopic theory}

We consider the system of $N$ bare
particles, each of them being one-dimensional massless
fermions, whose continuous energy spectrum is
$E(p)=cp$, with $c$ playing the role of speed of
light. We assume that these fermions are either `spinless''
(this means means that they all have the same direction of
spin and thus the spin degrees of fredom can be
neglected) or the 1+1 Dirac fermions.  If the fermions
are not interacting  the microscopic theory is extremely
simple: in vacuum state   fermions simply occupy all the
energy levels below the chemical potential
$\mu$.   In the continuous limit, the total number of
particles $N$ and the total energy of the system in the
one-dimensional ``cavity'' of size
$d$ are expressed in terms of the Fermi momentum
$p_F=\mu/c$ in the following way
\begin{eqnarray}
N=nd~,~n=\int_{-p_F}^ {p_F}{dp\over 2\pi\hbar} ={ p_F\over
\pi\hbar}~,
\label{FermiMomentum1}\\
E=\epsilon(n) d~,~\epsilon(n)=\int_{-p_F}^ {p_F}{dp\over
2\pi\hbar}cp  ={cp_F^2\over 2\pi\hbar}={\pi\over 2}
\hbar c  n^2~.
\label{FermiMomentum2}
\end{eqnarray}
Here  $\epsilon(n)$ is the  vacuum
energy density   as a function of the particle
density. The relation between the particle density and
chemical potential
$\mu=\pi
\hbar c n=p_F c$  also follows from
minimization of the relevant vacuum energy:
$d(\epsilon(n) -\mu n)/dn=0$.
In the vacuum state the relevant vacuum energy density and
the pressure of the Fermi gas  are
\begin{equation}
\tilde\epsilon=\epsilon(n) -\mu n=
-{\pi\over 2}  \hbar c  n^2~,~P=-\tilde\epsilon={\pi\over
2}  \hbar c  n^2~.
\label{FermionicVacuumEnergy}
\end{equation}
Fermi gas can exist only at positive external
pressure provided by the walls.

\subsection{Vacuum energy in effective theory}

As distinct from the microscopic theory, which deals with bare particles,
the effective theory deals with the  quasiparticles -- fermions living at the
level  of the chemical potential $\mu=cp_F$. There are 4 different
quasiparticles: (i)  quasiparticles and quasiholes living
in the vicinity of the Fermi point at
$p=+p_F$ have spectrum $E_{qp}(p_+)=|E(p)
-\mu|=c|p_+|$, where $p=p_z-p_F$;   (ii) quasiparticles
and quasiholes living in the vicinity of the other Fermi
point at
$p=-p_F$ have the spectrum $E_{qp}(p_-)=|E(p)
-\mu|= c|p_-|$,  where $p_-=p+p_F$. In the effective
theory the energy of the system  is the energy of the Dirac
vacuum
\begin{equation}
E=-\sum_{p_+}c|p_+|  -\sum_{p_-}c|p_-| ~.
\label{FermionicQuasiVacuumEnergy}
\end{equation}
 This energy is
divergent and requires the cut-off.  With the proper
cut-off provided by the Fermi-momentum,   $p_{\rm
Planck} \sim p_F$, the negative vacuum energy density
$\epsilon(n)$ in  Eq.(\ref{FermionicVacuumEnergy}) can be
reproduced.   This is a rather rare situation when the
effective theory gives the correct sign of the vacuum
energy.

\subsection{Vacuum energy as a function of discrete $N$}

Now let us discuss the Casimir effect -- the change of the vacuum pressure
caused by the finite size effects in the vacuum.   We must
take into account the discreteness of the spectrum  of
bare particles or quasiparticles (depending on which theory
we use, microscopic or effective) in the slab. Let us start
with the microscopic description in terms of bare
particles (atoms). We can use two different boundary
conditions for particles,  which give two kinds of discrete
spectrum:
\begin{eqnarray}
~E_k=k{\hbar c\pi\over d}~,
\label{LinearSpectrum1}\\
~E_k=\left(k+{1\over 2}\right){\hbar c\pi\over d}~.
\label{LinearSpectrum2}
\end{eqnarray}
Eq.(\ref{LinearSpectrum1})  corresponds to the
spinless fermions with Dirichlet boundary
conditions at the walls, while Eq.(\ref{LinearSpectrum2})
describes the energy levels of the 1+1 Dirac fermions with
no particle current through the wall; the latter case with
the generalization to the
$d+1$ fermions has been discussed in \cite{Paola}.

The vacuum is again represented by the ground state of the
collection of the $N$ noninteracting particles. We know the
structure of the completely and thus the vacuum energy in
the slab is well defined: it is the energy of  $N$
fermions in 1D box of size $d$
\begin{eqnarray}
E(N,d)=\sum_{k=1}^N E_k={\hbar c\pi\over 2d}N(N+1)
~~,~{\rm for}~~E_k=k{\hbar c\pi\over d}~,
\label{TotalEnergy1}\\
E(N,d)=\sum_{k=0}^{N-1}E_k={\hbar c\pi\over 2d}N^2
~~,~~{\rm for}~~E_k=\left(k+{1\over 2}\right){\hbar
c\pi\over d}~.
\label{TotalEnergy21}
\end{eqnarray}

\subsection{Leakage of vacuum through the wall.}

To calculate  the Casimir force acting on the wall, we
must introduce the vacuum on both sides of the wall. Thus
let us consider three walls: at
$z=0$, $z=d_1<d$ and $z=d$. Then we have two slabs with
sizes
$d_1$ and $d_2=d-d_1$, and we can find the force acting on
the  wall separating the two slabs, i.e. on the wall at
$z=d_1$. We assume the same  boundary conditions
at all the walls.  But we must allow the exchange
the particles between the slabs, otherwise the
main force acting on the wall between the slabs will be
determined simply by the difference in bulk pressure in the
two slabs.  This can be done due to, say, a very small
holes (tunnel junctions) in the wall,  which do not
violate the boundary conditions and thus do not disturb the
particle energy levels, but still allow the particle
exchange between the two vacua.

This situation can be compared with  the traditional
Casimir effect. The force between the conducting plates
arises because the electromagnetic fluctuations of the
vacuum in the slab are modified due to boundary conditions
imposed on electric and magnetic fields. In reality these
boundary conditions are applicable only in the
low-frequency limit, while the wall is transparent for the
high-frequency electromagnetic modes, as well as for the
other degrees of freedom of real vacuum (fermionic and
bosonic), that can easily penetrate through the conducting
wall. In the traditional approach it is assumed that those
degrees of freedom, which produce the divergent terms in
the vacuum energy, must be cancelled by the proper
regularization scheme.  That is why, though the dispersion
of dielectic permeability does weaken the real Casimir
force, nevertheless in the limit of large distances,
$d_1\gg  c/\omega_0$, where
$\omega_0$ is the characteristic frequency at which the
dispersion becomes important, the Casimir force does not
depend on how easily the high-energy vacuum leaks through
the conducting wall.

We consider here just the opposite limit, when (almost) all
the bare particles are totally reflected. This corresponds
to the case when the penetration of the high-energy modes of
the vacuum through the conducting wall is highly
suppressed, and thus one must certainly have the
traditional Casimir force.  Nevertheless, we shall show
that due to the mesoscopic finite-$N$ effects the
contribution of the diverging terms to the Casimir effect
becomes dominating. They produce highly oscillating vacuum
pressure in quantum liquids. The amplitude of the
mesoscopic fluctuations of the vacuum pressure in this
limit exceeds by factor $p_{\rm Planck}d/\hbar$  the value
of the conventional Casimir pressure. For their description
the continuous effective low-energy theories are not
applicable.

\subsection{Mesoscopic Casimir force in 1d Fermi gas}

The total vacuum energy in two slabs for spinless and
Dirac fermions is correspondingly
\begin{eqnarray}
E(N,d_1,d_2)={\hbar c\pi\over 2}\left({N_1(N_1+1) \over
d_1}+{N_2(N_2+1)
\over d_2}\right)~,
\label{TotalEnergyTwoBoxes1}\\
E(N,d_1,d_2)={\hbar c\pi\over 2}\left({N_1^2 \over
d_1}+{N_2^2
\over d_2}\right)~,
\label{TotalEnergyTwoBoxes2}
\end{eqnarray}
where $N_1$ and $N_2$ are the particle numbers in each of
the two slabs:
\begin{equation}
N_1+N_2=N~,~d_1+d_2=d
\label{ParticleConservation}
\end{equation}
Since particles can transfer between the slabs, the global
vacuum state in this geometry is obtained by minimization
over the discrete particle number
$N_1$ at fixed total number $N$ of particles.
If the mesoscopic
$1/N$ corrections are ignored, one obtains
$N_1\approx(d_1/d)N$ and $N_2\approx (d_2/d)N$; the
two vacua have the same pressure, and thus there is no
force acting on the wall between the two vacua.

However, $N_1$ and $N_2$ are integer valued, and this leads
to mesoscopic fluctuations of the Casimir force. The
global vacuum with given values of $N_1$ and $N_2$ is
realized only within a certain range of parameter
$d_1$. If $d_1$ increases, it reaches some treshold value
above which the energy of the vacuum with the
particle numbers $N_1+1$ and
$N_2-1$ has lower energy and it becomes the global
vacuum. The same happens if  $d_1$ decreases and reaches
some treshold value below which the vacuum
with the particle numbers $N_1-1$ and
$N_2+1$  becomes the global
vacuum.   The force
acting on the wall in the state ($N_1$,
$N_2$) is
obtained by variation of $E(N_1,N_2,d_1,d-d_1)$ over
$d_1$ at fixed $N_1$ and $N_2$:
\begin{equation}
F(N_1,N_2,d_1,d_2)=-{dE(N_1,N_2,d_1,d_2)\over
dd_1}+{dE(N_1,N_2,d_1,d_2)\over dd_2}~.
\label{TotalForce}
\end{equation}
When $d_1$ increases reaches the treshold, where
$E(N_1,N_2,d_1,d_2)=E(N_1+1,N_2-1,d_1,d_2)$, one particle
must cross the wall from the right to the left. At this
critical value the force acting on the wall changes
abruptly (we do not discuss here an interesting physics
arising just at the critical values of
$a_1$, where the degeneracy occurs between the states
($N_1,N_2$)  and  ($N_1+ 1,N_2- 1$); at these positions of
the wall (or membrane) the particle numbers
$N_1$ and
$N_2$ are undetermined and are actually fractional due to
the quantum tunneling between the slabs
\cite{Andreev}). Using for example the spectrum in
Eq.(\ref{TotalEnergyTwoBoxes2}) one obtains for the jump of
the Casimir force:
\begin{equation}
F(N_1\pm 1,N_2\mp 1)-F(N_1,N_2)  =\hbar c\pi\left({\pm
2N_1 +1\over 2d_1^2}+{\pm 2N_2-1 \over
2d_2^2}\right)\approx \pm {\hbar c\pi N\over d_1d_2} ~.
\label{ChaoticForceChange}
\end{equation}
The same result for the amplitude of
the mesoscopic fluctuations is obtained if one uses the spectrum in
Eq.(\ref{TotalEnergyTwoBoxes1}).

In the limit
$d_1\ll d$ the amplitude of the mesoscopic Casimir force
\begin{equation}
|\Delta F_{\rm meso}|= {\hbar c\pi n\over d_1}={\hbar
c\pi n^2\over N_1}\equiv { E_{Planck}\over d_1}~.
\label{AmplitudeForceChange}
\end{equation}
  It is  by factor $1/N_1=
(\pi\hbar/d_1p_{F})^3\equiv  (\pi\hbar/d_1p_{Planck})^3$
smaller than the vacuum energy density in
Eq.(\ref{FermiMomentum2}). On the other hand it is by the
factor
$p_Fd_1\equiv p_{Planck}d_1$ larger than the
traditional Casimir pressure, which in one-dimensional case
is $P_{\rm C}\sim  \hbar c/ d_1^2$. The divergent term
which linearly depends on the Planck momentum cutoff
$p_{Planck}$ as in Eq.(\ref{AmplitudeForceChange}) has
been revealed in many different calculations (see e.g.
\cite{CasimirForSphere}),  and attempts have made to invent
the regularization scheme which would cancel the divergent
contribution.

\subsection{Mesoscopic Casimir pressure in quantum liquids}

The equation (\ref{AmplitudeForceChange}) for the
amplitude of the mesoscopic fluctuations of the vacuum
pressure can be immediately  generalized for the
$d$-dimensional space: if
$V_1$ is the volume of the internal region separated by almost
inpenetrable walls from the outside vacuum, then the amplitude of the
mesoscopic vacuum pressure must be of order
\begin{equation}
|P_{\rm meso}|\sim {E_{Planck}\over V_1}~.
\label{MesoscopicVacuumPressure}
\end{equation}
The mesoscopic random pressure comes from the discrete
nature of the underlying quantum lquid, which represents
the quantum vacuum. The integer value of the number of
atoms in the liquid leads to the mesoscopic fluctuations of
the pressure: when the volume $V_1$ of the vessel changes
continuously, the equilibium number $N_1$ of particles
changes in step-wise manner. This results in abrupt changes
of pressure at some critical values of the volume:
\begin{equation}
P_{\rm meso}\sim P(N_1\pm 1)-P(N_1)=\pm {dP\over dN_1}
=\pm {mc^2\over V_1}\equiv \pm { E_{Planck}\over V_1}~,
\label{ChaoticForceChangeGeneral}
\end{equation}
where again $c$ is the speed of sound, which plays the
role of the speed of light.  The mesoscopic pressure is
determined by microscopic ``transPlanckian'' physics, and
thus such microscopic quantity as the mass $m$ of the atom,
the ``Planck mass'', enters this force.

For the spherical shell of radius $R$
immersed in the quantum liquid the mesoscopic
pressure is
\begin{equation}
P_{\rm meso}\sim  \pm {mc^2\over
 R^3}\equiv
\pm  \sqrt{-g}E_{\rm Planck}\left({\hbar
c\over  R}\right)^3  ~.
\label{ChaoticForceChangeGeneralSpherical}
\end{equation}

\subsection{Mesoscopic vacuum pressure vs
conventional Casimir effect.}

Let us compare the mesoscopic vacuum pressure in
Eq.(\ref{ChaoticForceChangeGeneralSpherical}) with the
traditional Casimir pressure obtained within the effective
theory for the same spherical shell geometry. In
the effective theory (such as electromagnetic theory in
case of the original Casimir effect, and the
low-ferquency quantum hydrodynamics in quantum liquids) the
Casimir pressure comes from the bosonic and fermionic
low-energy modes of the system (electromagnetic modes in
the original Casimir effect or quanta of sound waves in
quantum liquids). In superfluids, in addition to phonons
the other low-energy sound-like collective are possible,
such as spin waves. These collective modes with linear
(``relativistic'') spectrum in quantum liquids play the role
of the relativistic massless scalar field. They obey
typically the Neumann  boundary conditions,
corresponding to the (almost) vanishing mass or spin
current through the wall (almost, because the vacua
inside and outside the shell must be connected).

If we believe in the
traditional  regularization schemes which cancel out the
divergent terms, the effective theory gives the Casimir
pressure for the spherical shell is
\begin{equation}
P_{\rm C}=-{dE_{\rm C}\over dV}={K
\over 8\pi}\sqrt{-g}\left({\hbar c\over  R}\right)^4 ~,
\label{TradCasSpherical}
\end{equation}
where $K=-0.4439$ for the Neumann boundary conditions;
$K=0.005639$ for the Dirichlet boundary conditions
\cite{CasimirForSphere}; and $c$ is the speed of
sound or of spin waves. The traditional Casimir pressure is
completely determined by the effective low-energy theory,
it does not depend on the microscopic structure of the
liquid: only the ``speed of light'' $c$ enters this
force.  The same pressure will be obtained in case of the
pair correlated fermionic superflids, if the fermionic
quasiparticles are gapped and their contribution to the
Casimir pressure is exponentially small compared to the
contribution of the collective massless bosonic modes.

However, at least in our case, the result obtained within
the effective theory is not correct: the real Casimir
pressure is given by
Eq.(\ref{ChaoticForceChangeGeneralSpherical}): (i) It
 essentially depends on the Planck
cut-off parameter, i.e. it cannot be determined by the
effective theory; (ii) it is much bigger, by factor
$p_{\rm Planck}R/\hbar$, than the traditional
Casimir pressure in Eq.(\ref{TradCasSpherical});  and (iii)
it is highly oscillating. The regularization of these
oscillations by, say, averaging over many measurements; by
noise; or due to quantum or thermal fluctuations of the
shell; etc., depend on the concrete physical conditions of
the experiment.

This shows that in some cases the
Casimir vacuum pressure is not within the responsibility
of the effective theory, and the microscopic
(trans-Planckian) physics must be evoked. If two systems
have the same low-energy behavior and are described
by the same effective theory, this does not
mean that they necessarily experience the same Casimir
effect. The result depends on many factors, such as the
discrete nature of the quantum vacuum, and the ability of
the vacuum to penetrate through the boundaries. It is not
excluded that even the traditional Casimir effect which
comes from the vacuum fluctuations of the electromagnetic
field is renormalized by the high-energy degrees of freedom

Of course, the extreme limit, which we consider, is not
applicable to the original (electromagnetic)  Casimir
effect, since the situation in the electromagnetic Casimir
effect is just opposite. The overwhelming part of the
fermionic and bosonic vacuum easily penetrates the
conducting wall, and thus the mesoscopic fluctuations are
small. But do they negligibly small? In any case this
example  shows that the cut-off problem is not the
mathematical, but the physical one, and the Planck physics
dictates the proper regularization scheme or the proper
choice of the cut-off parameters.

\section{Conclusion.}

We discussed the problems related to the properties of
quantum vacuum in general relativity using the known
properties of the quantum vacuum in quantum liquids, where
some elements of the Einstein gravity arise in the
low-energy corner. We found that in both systems there are
similar   problems, which arise if the effective theory is
exploited. In both systems the naive estimation of the
vacuum energy density within the effective theory gives
$\rho_{\Lambda}\sim E_{\rm Planck}^4$ with the
corresponding ``Planck'' energy appropriate for each
of the two systems. However, as distinct from the general
relativity, in quantum liquids the fundamental physics,
``The Theory of Everything'', is known, and it
shows that the ``trans-Planckian'' degrees of freedom
exactly cancel this divergent contribution to the vacuum
energy. The relevant vacuum energy is zero without fine
tuning, if the vacuum is stable (or metastable), isolated
and homogeneous.

Quantum liquids also demonstrate how the small vacuum
energy is generated, if the vacuum is disturbed. In
particular, thermal quasiparticles -- which represent the
matter in general relativity -- induce the vacuum energy
of the order of the energy of the matter. This example
shows the possible answer to the question, why the present
cosmological constant is of the order of the present matter
density in our Universe.  It follows that in each epoch the
vacuum energy density must be of order of either the matter
density of the Universe, or of its curvature, or
of the energy density of the smooth component -- the
quintessence. However, the complete understanding of the
dynamics of the vacuum energy in the time-dependent regime
of the expanding Universe cannot be achieved within
the general relativity and requires the extension of this
effective theory.

In principle, one can construct the artificial quantum
liquid, in which all the elements of the general
relativity are reproduced in the low energy corner. The
effective metric $g^{\mu\nu}$ acting on ``relativistic''
quasiparticles arises as one of the low-energy collective
variables of the quantum vacuum, while the Sakharov
mechanism leads to the Einstein curvature and cosmological
terms in the action for this dynamical variable. In this
liquid the low energy phenomena will obey the Einstein
equations (\ref{EinsteinEquation}), with probably one
exception: the dynamics of the cosmological ``constant''
will be included. It would be extremely interesting to
realize this programme, and thus to find out the possible
extension of general relativity, which takes into account
the properties of the quantum vacuum.

The most important property of the quantum vacuum in
quantum liquids is that this vacuum consists of
discrete elements -- bare atoms. The interaction and zero
point oscillations of these elements lead to the formation
of the equilibrium vacuum, and in this equilibtium vacuum
state the cosmological constant is identically zero.
Thus the discreteness of the quantum vacuum can be the
possible source of the (almost complete) nullification of
the cosmological constant in our present Universe. If so,
one can try to exploit the other possible consequences of
the discrete nature of the quantum vacuum, such as the
mesoscopic Casimir effect discussed in Sec.
\ref{EffectsDiscreteNumberN}.

Analogy with the quantum vacuum in quantum liquids allows
us to discuss the other problems related to the quantum
vacuum in general relativity: the flatness problem; the
problem of a big entropy in the present Universe; the
horizon problem, etc.

 This
work  was supported in part by the Russian Foundation for
Fundamental Research and by European Science Foundation.

\end{document}